\documentclass[lettersize,journal]{IEEEtran}
\usepackage{amsmath,amsfonts}
\usepackage{algorithmic}
\usepackage[ruled,vlined,linesnumbered]{algorithm2e}
\usepackage{array}
\usepackage[caption=false,subrefformat=parens,labelformat=simple]{subfig}
\usepackage{textcomp}
\usepackage{stfloats}
\usepackage{url}
\usepackage{verbatim}
\usepackage{graphicx}
\usepackage{cite}
\usepackage{comment}
\usepackage{multirow}

\newtheorem{theorem}{Theorem}
\newtheorem{definition}{Definition}
\newtheorem{lemma}{Lemma}
\newtheorem{corollary}{Corollary}
\begin{document}

\title{Optimization Framework for Splitting DNN Inference Jobs over Computing Networks}

\author{Sehun~Jung,
        and~Hyang-Won~Lee,~\IEEEmembership{Member,~IEEE}% <-this % stops a space
\thanks{The authors are with the Department of Computer Science and Engineering, Konkuk University, Seoul, Republic of Korea.}%
% note need leading \protect in front of \\ to get a newline within \thanks as
% \\ is fragile and will error, could use \hfil\break instead.
\thanks{e-mail: \{qhsl1213, leehw\}@konkuk.ac.kr}
\thanks{This work was supported by the National Research Foundation of Korea (NRF) grant funded by the Korea government (MSIT) (No.2021R1A2C2012801). (Corresponding author: Hyang-Won Lee)}% <-this % stops an unwanted space
%\thanks{Manuscript received April 19, 2005; revised August 26, 2015.}
}

%\author{IEEE Publication Technology,~\IEEEmembership{Staff,~IEEE,}
%        % <-this % stops a space
%\thanks{This paper was produced by the IEEE Publication Technology Group. They are in Piscataway, NJ.}% <-this % stops a space
%\thanks{Manuscript received April 19, 2021; revised August 16, 2021.}}

% The paper headers
%\markboth{IEEE Internet of Things Journal,~Vol.~XX, No.~X, Month~2022}%
%{Shell \MakeLowercase{\textit{et al.}}: A Sample Article Using IEEEtran.cls for IEEE Journals}

%\IEEEpubid{0000--0000/00\$00.00~\copyright~2022 IEEE}
% Remember, if you use this you must call \IEEEpubidadjcol in the second
% column for its text to clear the IEEEpubid mark.

\maketitle

\begin{abstract}
Ubiquitous artificial intelligence (AI) is considered one of the key services in 6G systems. AI services typically rely on deep neural network (DNN) requiring heavy computation. Hence, in order to support ubiquitous AI, it is crucial to provide a solution for offloading or distributing computational burden due to DNN, especially at end devices with limited resources. We develop an optimization framework for assigning the computation tasks of DNN inference jobs to computing resources in the network, so as to reduce the inference latency. To this end, we propose a layered graph model with which simple conventional routing jointly solves the problem of selecting nodes for computation and paths for data transfer between nodes. We show that using our model, the existing approaches to splitting DNN inference jobs can be equivalently reformulated as a routing problem that possesses better numerical properties. We also apply the proposed framework to derive algorithms for minimizing the end-to-end inference latency. We show through numerical evaluations that our new formulation can find a solution for DNN inference job distribution much faster than the existing formulation, and that our algorithms can select computing nodes and data paths adaptively to the computational attributes of given DNN inference jobs, so as to reduce the end-to-end latency.
\end{abstract}

\begin{IEEEkeywords}
DNN job splitting, computing network, completion time, routing
\end{IEEEkeywords}

\section{Introduction}\label{sec:intro}
6G system is envisioned to support artificial intelligence (AI) services all over the network from the core to the end hosts, referred to as ubiquitous AI \cite{letaief:roadmap}. In many cases, AI services depend on the computation of deep neural network (DNN) and hence require a fair amount of computing power, even only for inference. This can be a significant burden especially for end devices such as mobile phones and IoT devices in which computing resources are highly limited as they run on limited battery power. In order for ubiquitous AI service to be in place, it is thus necessary to provide a solution to overcome limited computing power. 

There are several approaches to enable AI at end devices. One is to find a lightweight NN architecture commensurate with available computing resources. SqueezeNet in \cite{DBLP:journals/corr/IandolaMAHDK16} makes extensive use of 1x1 convolutions to reduce the number of parameters. MobileNetV1 in \cite{DBLP:journals/corr/HowardZCKWWAA17} reduces the number of arithmetic operations by introducing depth-wise separable convolutions. In \cite{howard:mobilenetv3}, automated neural architecture search (NAS) is proposed based on reinforcement learning in which the reward reflects the latency of inference. Since the latency depends on the underlying computing resources, such a reward  drives the action (i.e., values of hyperparameters) toward the set of architectures that can yield reasonably fast inference for given computing power. There is a large body of work in this context, and refer to \cite{ren:survey} for more details.

Another approach is to exploit computing resources distributed over the network. Specifically, the (feedforward) computation of DNN inference job is partitioned into multiple tasks, and these computation tasks are assigned to nodes with computing resources in the physical network. For example, in layer-wise partition, all the neurons in the same layer are assigned to a same node \cite{baccour:RLPDNN, disabato:DDCNN}. Once the node finishes computing the tasks (or layers) assigned, it transfers the output data (of the last layer assigned) to the next node and subsequently, the transferred data are fed as the input data to the corresponding layer. In this work, we develop an optimization framework for distributing or splitting computation tasks of DNN inference jobs over the network in which some (or all) nodes are equipped with computing resources.

Such a problem requires to make a joint decision on node selection for computation and path selection for data transfer. One of the challenges in the problem is that the amount of flow can change after passing through a node where computation is carried out, because input and output data size of DNN layer can differ from each other. Hence, the problem is vastly different from conventional routing in which flow conservation holds, i.e., incoming and outgoing flows match at every node except at source and destination. To tackle the problem, we propose a layered graph model in which simple conventional routing jointly solves the node and path selection problems. Using the model, we reformulate the existing approaches to distributing DNN inference jobs as a conventional routing problem that possesses better numerical properties.

Furthermore, we apply the proposed framework for computing DNN inference jobs over the network with minimum end-to-end latency, defined as the duration between the time when the data at source start to be processed and the time when the inference result is delivered to the destination. Obviously, the end-to-end inference latency consists of waiting time and service time. The waiting occurs at link(s) when data need to wait to be transmitted, and also at node(s) when computation tasks need to wait to be processed. The service time is the pure transmission time plus computation time. With this definition, many of the existing works in the context of DNN inference job splitting focus primarily on minimizing the service time while it is also important to take into account the waiting time. %There are several challenges

It is generally hard to deal with waiting time as it is a complex function of arrival and departure processes. Nonetheless, we consider a fictitious system in which the waiting time is an upper bound on the waiting time in the actual system. Our framework enables to efficiently solve the problem of distributing a single DNN inference job such that the end-to-end latency is minimized in the fictitious system. Exploiting the efficient solvability of single-job problem with our framework, we develop algorithms for distributing multiple DNN inference jobs so as to reduce the job completion time defined as the earliest time at which all the inference jobs are finished. The contributions of our work can be summarized as follows:
\begin{itemize}
\item We develop an optimization framework for computing DNN inference jobs over distributed computing networks with minimum latency. Our framework enables to select nodes for computation and paths for data transfer jointly via simple conventional routing in the layered graph model.
\item Applying our framework, we reformulate the existing approaches as a conventional routing problem that exhibits superior numerical performance.
\item We develop algorithms for distributing DNN inference jobs over the network so that the end-to-end latency is reduced.
\item We prove the numerical property of our formulation and the performance guarantee of our algorithms.
\item We verify the performance of our formulations and algorithms through various numerical evaluations.
\end{itemize}

The rest of the paper is organized as follows. In Section \ref{sec:related-work}, we discuss related work. In Section \ref{sec:system-model}, we present the system model and describe the problem, and in Section \ref{sec:layered-graph-formulation}, we propose the layered graph model that simplifies the DNN job distribution as a conventional routing problem. In Section \ref{sec:applications}, we present some applications of our framework including 1) reformulation of existing approaches and 2) algorithms for computing DNN inference jobs with minimum completion time. In Section \ref{sec:numerical-evaluation}, we present numerical evaluations demonstrating the performance of our formulations and algorithms. We conclude the paper in Section \ref{sec:conclusion}.
%All the missing proofs can be found in \cite{extended-version}.

\section{Related Work}\label{sec:related-work}
%There is an increasing demand for end hosts to support compute-intensive services such as deep learning based image analysis. In many cases, these computations are hard to be carried out at end hosts which typically have limited computing resources. Offloading computing tasks 

%There are many solutions to overcoming the computational limit of end hosts. 
Edge computing is one of the most promising solutions in the context of overcoming the computational limit at end hosts \cite{shi:edge}. The demand for edge computing mainly arises from video or image analytics that heavily relies on compute-intensive deep learning \cite{wang:convergence}. In many cases, the focus is on how to select either the local computing resource or remote (edge or cloud) server, so that the entire analysis of an image or frame can be carried out at the chosen resource. For example, in \cite{liu:dare}, video frames are sent to edge server in which one of neural network based inference models is selected for object detection. In \cite{tan:deep}, the offloading decision of each video frame is made based on estimated network condition and response time, so that the end-to-end latency of frame analysis can be minimized. In \cite{wu:accuracy}, the offloading problem is solved taking into account service time and accuracy in the setting where the local device has a small model while the edge has cumbersome model with better accuracy.

As mentioned above, the demand for computation offloading arises mainly when inference or analysis should be made based on deep neural network (or deep learning). Since deep neural network typically consists of multiple layers and filters (applied block-by-block in data) in some early layers, there has been effort to distribute and parallelize the computation tasks of layers and/or blocks in layers, which is of main interest in this work. 

Many of works in this context consider the Inter-of-Things (IoT) environment in which nodes are equipped with limited computing resources. In \cite{pascale:network}, each hidden neuron is mapped to an IoT node. The goal is to find a mapping that minimizes the network-wide total transmit power or time while the computing resource needed for assigned neurons and daily energy consumption satisfy the specified limit. The work of \cite{baccour:RLPDNN} assumes layer-wise partition, and seeks to find a mapping that minimizes (data) transmission time plus computation time, subject to privacy constraints as well as various resources constraints. This problem is formulated as an integer quadratic program, and a reinforcement learning based mapping algorithm is proposed. In \cite{disabato:DDCNN}, a similar problem and formulation are developed with early-exit convolutional neural network (CNN).

Some work considers vertical partition of convolution operations that take the majority of computation in DNN. Unlike the layer-wise partition where partitioned tasks have precedence constraint, in vertical partition, convolution operations are partitioned into tasks that can be computed independently. In \cite{zhao:deepthings}, an IoT node distributes such independent computation tasks (formed through vertical partition) to nearby IoT devices and fuses the collected results at the output of convolution layer. In \cite{mao:modnn, hadidi:toward}, even a layer in the fully connected part is partitioned into multiple groups and distributed to computing nodes. Since parallelization inevitably incurs communication overhead, some works seek to find a partition and allocation of computation tasks with minimum communication demand \cite{stahl:fully,xue:edgeld}. There are also some work considering both vertical and layer-wise partitions together with model sparsification \cite{chang:EDDL}.

While many of the above solutions are derived based on the formulation of integer optimization \cite{disabato:DDCNN,pascale:network,stahl:fully,he:joint}, there are many results that apply deep reinforcement learning (DRL), circumventing the difficulty of solving such an optimization problem (which is NP-hard in general) \cite{baccour:RLPDNN}. In \cite{zhang:learning}, the classical job-shop scheduling problem is solved via DRL adopting graph neural network, in which a job consists of sequential operations with precedence constraint. This is related to the distributed DNN computation as layers in DNN are computed sequentially with precedence. In \cite{huang:toward}, a computation task is formed with a group of neuron(s) from a single layer, and the RL agent distributes the tasks to mobile devices with reward being aggregate computation throughput, which has the effect of minimizing latency. This work is extended to account for partition of convolution operations  \cite{huang:enabling}. In \cite{xue:ddpqn}, DRL-based algorithm makes a decision on each layer of DNN inference computation whether the layer is computed at local, edge, or cloud, so as to minimize the integrated objective of delay, energy and cost. \cite{xu:energy} develops an RL-based algorithm for deferring the assignment of incoming inference request, leaving a room for future requests, which enables a better packing of requests with smaller latency and energy consumption.
%\cite{bao:deep} %computation jobs from training

The goal of our work is focused on optimization framework for splitting DNN inference jobs over distributed computing networks. In particular, we are primarily interested in developing an efficient formulation of the problem. We believe that such a formulation can be utilized in various facets of methodology to solve the DNN inference job distribution problem.
%
%routing/ shortest path with link capacity constraints
%
%latency including 
%
%terminology: routing, path selection, node selection, computation flow routing

%\begin{itemize}
%%\item offloading (general methods)
%\item specific to NN distribution [6]-[10]
%\item new references: splitting types, optimization-based, RL-based
%\end{itemize}
%This can be viewed as a generalization of edge or cloud computing.
%
%For example, in 5G, edge computing is the de facto service considered my many 
%
%distribute computation tasks in DNN (inference in particular) and 
%
%movement toward convergence of communication and computation
%
%We focus on ...
%
%such as Google Lens

%\begin{itemize}
%%\item ubiquitous AI
%\item two approaches: lightweight neural architecture search, computational solution (distributed/parallel)
%\item NAS: MobileNet1,2,3, RL-based, ...
%\item distributed computation: vertical/horizontal(layer-wise) partition and distribute
%%\item layer-wise partition
%\end{itemize}

%\cite{pascale:network}, \cite{zhao:deepthings}, \cite{chang:EDDL}, \cite{baccour:RLPDNN}, \cite{disabato:DDCNN}

%\section{Related Work}
%\begin{itemize}
%\item budget constraints not make sense
%\end{itemize}

\section{System Model and Problem Description}\label{sec:system-model}
Consider a communication network in which nodes are equipped with computing resources. Let $G_P=(V_P,E_P)$ denote this physical network where $V_P$ is the set of nodes (routers/servers/hosts) and $E_P$ is the set of edges (communication links) connecting the nodes. Let $\mu_{uv}$ be the transmission capacity of link $(u,v)\in E_P$. The computation capacity of node $u$ is denoted as $\mu_u$, and its unit can for example be GFLOPs/sec. There is a queue for every transmission link, and $Q_{uv}$ is the queue length at link $(u,v)$ representing the amount of packets waiting to be transmitted. Likewise, $Q_u$ is the amount of computation tasks (e.g., in GFLOPs) waiting to be computed at node $u$. This computing network is used to process deep neural network (DNN) inference jobs.

\subsection{Inference Jobs}
There are $J$ DNN inference jobs, each corresponding to the feedforward computation of a DNN model (We call job and model interchangeably depending on the context). For each model $j\in\mathcal{J}=\{1,...,J\}$, the input data (e.g., camera images and sensor values) are generated at $s^j\in V_P$ and the inference result needs to be delivered to $t^j\in V_P$.  Each model $j\in\mathcal{J}$ has $L^j$ layers. We assume that each layer can be possibly computed at different nodes in $V_P$. Let $c_l^j$ be the computational load of layer $l(=1,...,L^j)$ of model $j$. Hence, $\frac{c^j_l}{\mu_u}$ is the \emph{computation time} if layer $l$ of model $j$ is to be processed at node $u$. Let $d^j_l$ be the output data size of layer $l$ of model $j$. Similarly, $\frac{d^j_l}{\mu_{uv}}$ is the \emph{transmission time} if the output data of layer $l$ is to be transferred from node $u$ to $v$. The computation time plus transmission time is called the \emph{service time}. Hence, if layer $l$ of model $j$ is computed at node $u$, and the output data is transferred to node $v$, then the service time at this segment is $\frac{c^j_l}{\mu_u} + \frac{d^j_l}{\mu_{uv}}$.

We also consider the \emph{waiting time}. In the above example, suppose that the queue length at node $u$ is $Q_u$ when the computation task arrives at node $u$, and the queue length at link $(u,v)$ is $Q_{uv}$ when the output of computation is buffered at link $(u,v)$ for transmission. Thus, the waiting time is $\frac{Q_u}{\mu_u} + \frac{Q_{uv}}{\mu_{uv}}$. The duration between the time of entering node $u$ and the time of arriving at node $v$ is waiting time plus service time, i.e., $\frac{Q_u}{\mu_u} + \frac{Q_{uv}}{\mu_{uv}} + \frac{c^j_l}{\mu_u} + \frac{d^j_l}{\mu_{uv}}$. In this work, for simplicity of presentation, we ignore propagation time at any component either inside a node or between nodes (i.e., link), but our results can be readily applied to the scenario with propagation delay since propagation adds constant delay.

\subsection{Job Completion Time}
Suppose that each job $j$ is assigned a path from its source $s^j$ to destination $t^j$, including the information of the node at which each layer is computed. Time starts from $0$. Let $C^j$ be the time when the inference result of job $j$ is delivered to the destination, i.e., $t^j$. Hence, $C^j$ is the end-to-end inference latency. This time is obviously equal to waiting time plus service time along the path from $s^j$ to $t^j$. Define the \emph{job completion time} $C_{\max}$ as
\begin{align*}
C_{\max} = \max_j C^j.
\end{align*}
Hence, $C_{\max}$ is the time when all the jobs are finished. This time is also called \emph{makespan} in the field of job-shop scheduling. Clearly, the job completion time is determined by path selection for data transfer and node selection for computation. We simply call these two decisions ``routing''. \emph{In this work, we develop a framework for routing DNN inference jobs for minimum job completion time.}

\subsection{Challenges}
There are several challenges in tackling the problem of routing DNN inference jobs for minimum completion time. First, the routing problem in this work is inherently related to the classical job-shop scheduling problem which is known to be NP-complete \cite{garey:computers}. In the job-shop scheduling problem, there are multiple jobs, and each job consists of ordered operations. All of the operations must be assigned to some machine, and in each machine, the priority among operations should be decided in order to minimize the job completion time. If the machine for each operation to be computed at is given and fixed, then the problem is called the ``job-shop scheduling problem''. Otherwise, it is called the ``flexible job-shop scheduling problem''. In our problem, each layer (corresponding to operation in job-shop scheduling) in every job should be assigned a node for computation, together with path selection. Our problem therefore contains a flexible job-shop scheduling which is hard to solve. Note that this is immensely different from the conventional routing problem which is easy to solve under many circumstances. 
%We develop heuristic algorithms based on the layered graph representation of the problem.

%Given a path selection from source to destination, the routing problem is exactly the same as flexible job-shop scheduling
%
%Given a node selection for computation (i.e., each layer in the model is assigned a computing node), 

Another challenge comes from the difficulty of handling waiting time. It is necessary to predict the waiting time so that the routing decision can be made by taking into account both waiting time and service time. However, the waiting time at a component (i.e., node or link) depends on the departure processes in the preceding components and hence, it is hard to predict the waiting time.
%\footnote{In contrast, measuring the waiting time is fairly easy.}. 
To circumvent this issue, we consider a fictitious system in which the waiting time provides an upper bound on the actual waiting time. The completion time in the fictitious system is thus an upper bound on the actual completion time. We apply our framework in order to find a routing decision minimizing completion time in the fictitious system. The details are presented in Section \ref{sec:applications}.

%Routing
%
%Given computing nodes, shortest path problem
%Given paths, job-shop scheduling problem
%
%Waiting time
%

\section{Layered Graph and Routing}\label{sec:layered-graph-formulation}
\begin{figure}[htbp]
\centerline{\includegraphics[scale=0.37]{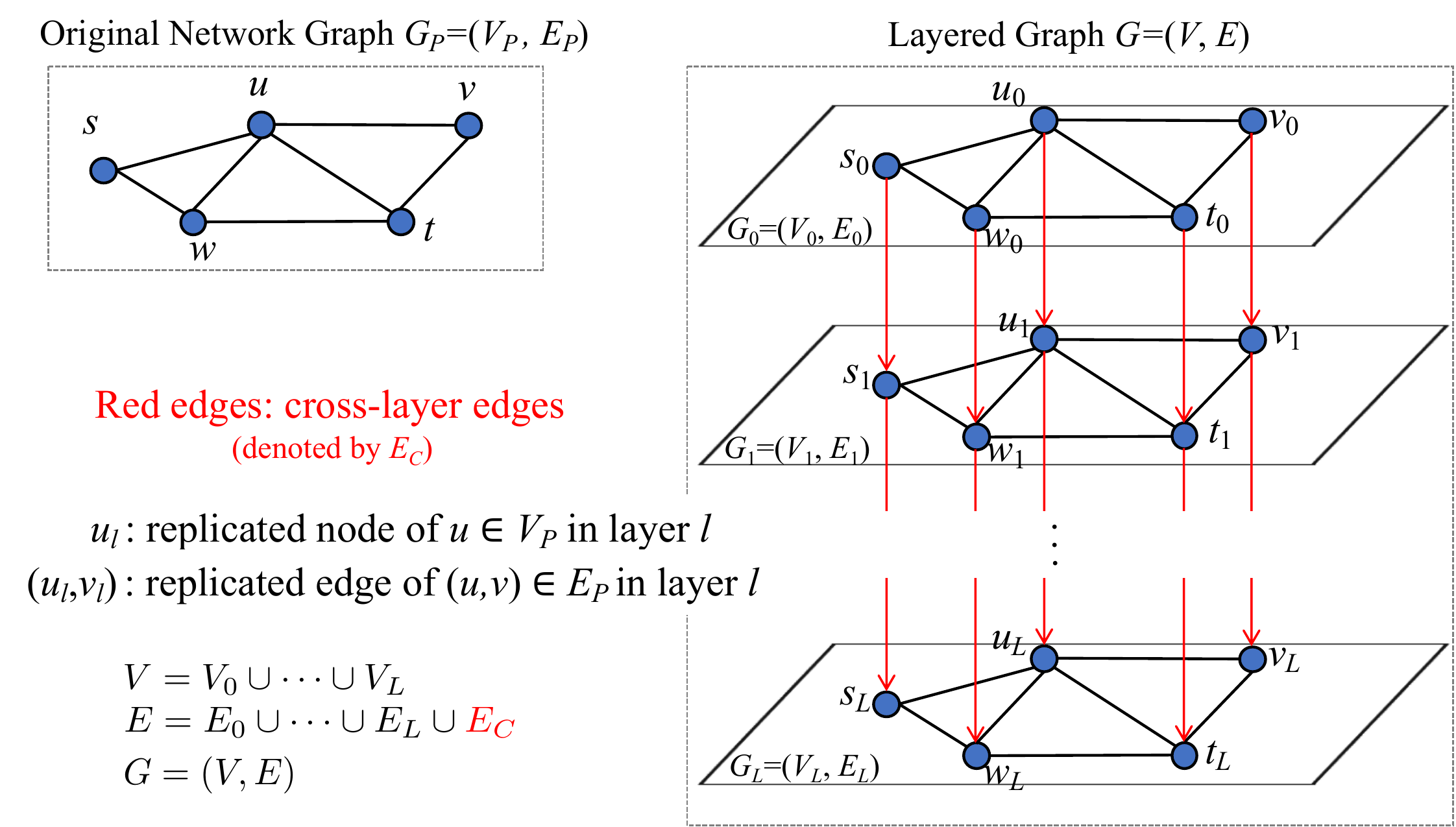}}
\vspace{-0.2cm}
\caption{Example of layered graph derived from original network graph $G_P$ for DNN model with $L$ layers}
\label{fig:layered-graph-example}
\vspace{-0.3cm}
\end{figure}

As mentioned above, our problem requires to determine the path as well as the nodes for computation along the path. It is hard to formulate this problem applying the traditional technique involving flow conservation constraints. Inspired by \cite{cao:enhancing} and \cite{zhang:optimal} where service chaining problem in software-defined network (SDN) is solved with graph layering, we construct the layered graph with which the problem can be simplified. Suppose for simplicity of presentation that there is only a single model with $L$ layers. Consider $L+1$ copies of $G_P$, denoted by $G_0$, $G_1$, ..., $G_L$ with $G_l=(V_l,E_l),\forall l$. For each $l\in\mathcal{L}=\{0,...,L\}$, denote by $u_l\in V_l$ the replicated node of $u\in V_P$, and hence, $(u_l,v_l)\in V_l$ the replicated link of $(u,v)\in E_P$. There is an edge from node $u_{l-1}$ to $u_l$ for all $u\in V_P$ and $l=1,...,L$. These edges are called \emph{cross-layer edges}, denoted by $E_C$. Define the layered graph $G=(V,E)$ where $V = V_0 \cup \cdots \cup V_L$ and $E = E_0 \cup \cdots \cup E_L \cup E_C$. Fig.~\ref{fig:layered-graph-example} shows an example of the layered graph $G$ derived from the original physical network graph $G_P$.

\subsection{Routing in Layered Graph}
\begin{figure}[htbp]
\centerline{\includegraphics[scale=0.4]{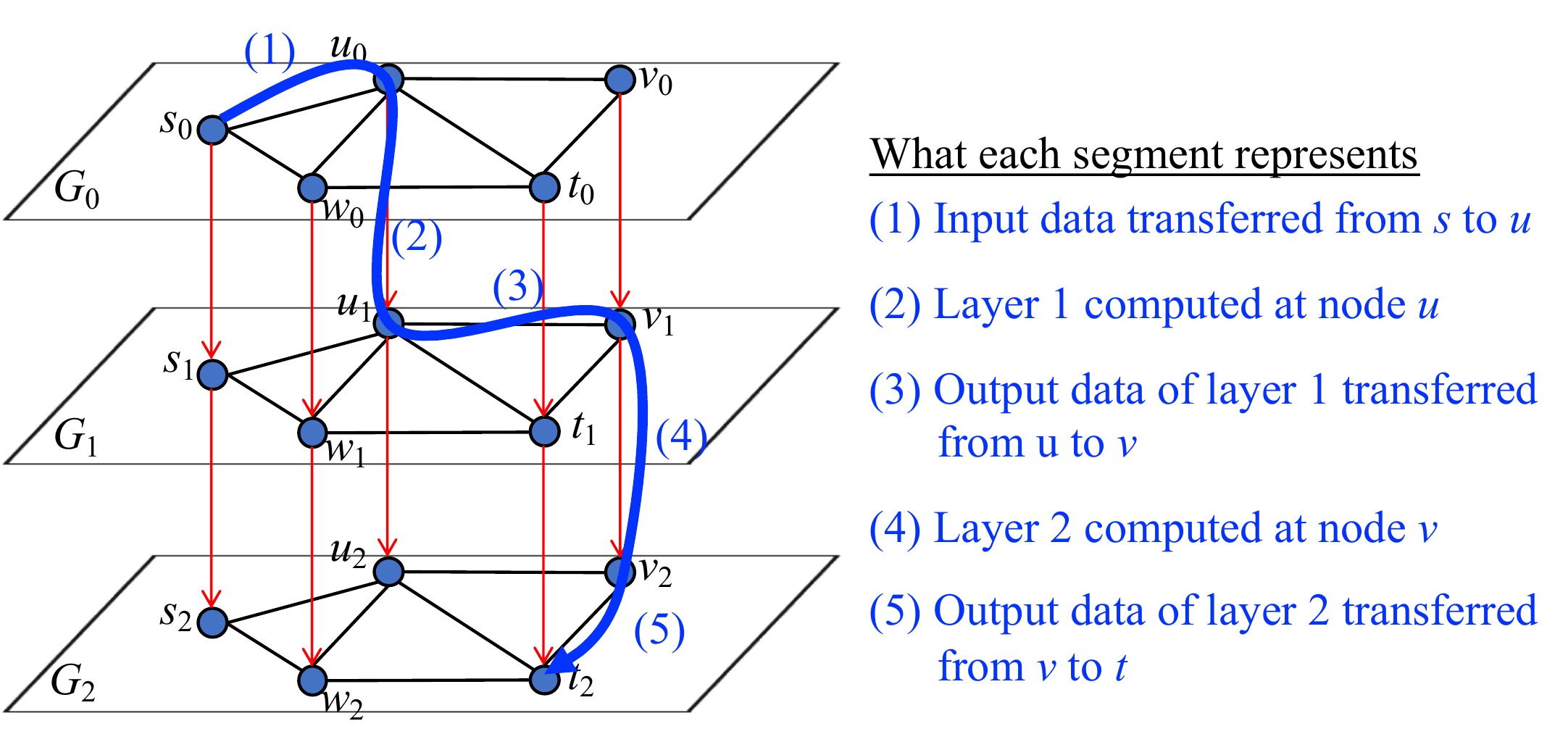}}
\vspace{-0.3cm}
\caption{Example of routing in layered graph with $L=2$}
\label{fig:routing-example}
\vspace{-0.3cm}
\end{figure}

We now discuss how the routing in the layered graph simply expresses both path selection for data transfer and node selection for computation. Suppose that source and destination nodes are $s\in V_P$ and $t\in V_P$ respectively. Consider finding a path from $s_0$ to $t_L$. The cross-layering segment of the path specifies the node where the corresponding layer is computed. For instance, if the path traverses link $(u_{l-1},u_l)$, then layer $l$ of the model is computed at node $u$. The intra-layer segment of the path specifies the transfer of the output data of corresponding layer (of model). For instance, if the path traverses the link $(u_l,v_l)$, then the output data of layer $l$ is transferred from node $u$ to node $v$. Fig.~\ref{fig:routing-example} shows an example of routing in the layered graph, and what each segment in the path represents. It is important to note that the classical routing (that just finds a path from source to destination) in the layered graph determines both path selection and node selection for computing of all the layers of the model simultaneously.

\subsection{Routing Multiple Jobs}
The layered graph can easily facilitate the routing of multiple jobs at the same time. Recall that $L^j$ and $(s^j, t^j)$ represent the number of layers in job (or model) $j$ and source-destination pair of job $j$, respectively. Redefine $L=\max_j L^j$, and construct the layered graph with $L+1$ layers as shown in Fig.~\ref{fig:layered-graph-example}. The source $s^j\in V_P$ and destination $t^j\in V_P$ are mapped to $s^j_0\in V$ and $t^j_{L^j} \in V$, respectively. Finding a path from $s^j_0$ to $t^j_{L^j}$ for each job $j$ in the layered graph gives the node and path selection of all the jobs. Some examples pertaining to this are presented in Section~\ref{sec:applications}.

%\subsection{Various Types of DNN}
%skip connection, AssembleNet-type DNN

\section{Applications of Our Framework}\label{sec:applications}
Our framework in the previous section can be applied to various problems of distributing DNN inference jobs. In this section, we present some of these examples.

\subsection{Reformulation of Existing Approaches}
The work of \cite{baccour:RLPDNN}  and \cite{disabato:DDCNN} formulates the problem of assigning DNN layers to nodes for minimizing end-to-end latency as an integer quadratic program (IQP). The data transmission time between two nodes, say $u$ and $v$, is taken into account by assuming that every link has the same capacity and the data are transferred on the shortest hop path from $u$ to $v$. Although this assumption of data rate simplifies the formulation, the end-to-end latency under the assignment obtained from such a formulation can experience relative large latency when data transmission time contributes to end-to-end latency to a large portion and links can have different capacities. Obviously, such a case occurs when networking resources are scarce while computing resources are abundant.

Our framework can incorporate both node and path selection as an integer linear program (ILP) which has numerical advantages over IQP. To unify the notation in layered graph, let us define the edge capacity and transmission/computation task as follows:
\begin{enumerate}
\item $\mu_{u_l v_l} = \mu_{uv},\forall (u,v)\in E_P, l\in\mathcal{L}$
\item $\mu_{u_{l-1} u_l} = \mu_u,\forall u\in V_P, l=1,...,L$
\item $q_{uv}^j = \begin{cases}
d_l^j,& \mbox{if } (u,v)\in E_l, l\in\mathcal{L}\\
c_l^j,& \mbox{if } (u,v)=(w_{l-1},w_l)\in E_C, l=1,...,L \end{cases}$,
\end{enumerate}
where $\mathcal{L}=\{0,1,...,L\}$. The first line indicates that every intra-layer edge $(u_l,v_l)\in E_l,\forall l\in\mathcal{L}$ in the layered graph has the same capacity as the original edge $(u,v)\in E_P$. The second line indicates that the capacity $\mu_{u_{l-1} u_l}$ of cross-layer edge $(u_{l-1},u_l),\forall l\in\mathcal{L}$ is equal to the computation rate of node $u$ which is $\mu_u$. The value $q_{uv}^j$ denotes the amount of task. If $(u,v)$ is the intra-layer edge in $G_l$, then $q^j_{uv}$ is the data size of layer $l$ of model $j$. On the other hand, if $(u,v)$ is the cross-layer edge from $w_{l-1}$ to $w_l$ for any $w$, then $q^j_{uv}$ is the amount of computation tasks required by layer $l$ of model $j$. Hence, $\frac{q^j_{uv}}{\mu_{uv}}$ is the transmission time of output data of layer $l$ if $(u,v)$ is an intra-layer edge in $E_l$, and the computation time of layer $l$ at node $w$ if $(u,v)$ is a cross-layer edge $(w_{l-1},w_l)$.

Define the variable $r_{uv}^j$ to be 1 if the path of job $j$ traverses edge $(u,v)\in E$, and 0 otherwise. Let $m_l^j$ be the memory requirement of layer $l(=1,...,L^j)$ of model $j$. Consider the following formulation:
\begin{align}
\min_r &\quad \sum_{j\in\mathcal{J}}\sum_{(u,v)\in E} \frac{q_{uv}^j}{\mu_{uv}} r_{uv}^{j}\label{eqn:obj-simdis}\\
\mbox{s.t.} & \quad \sum_{v:(u,v)\in E} r_{uv}^j -  \sum_{v:(v,u)\in E} r_{vu}^j = \begin{cases}
1,& \mbox{if } u = s_0^j\\
-1,& \mbox{if } u = t_{L^j}^j\\
0,& \mbox{otherwise}
\end{cases},\nonumber\\
&\qquad\forall u\in V, j\in\mathcal{J}\label{eqn:con-path}\\
&\quad \sum_{j\in\mathcal{J}}\sum_{l=1}^{L^j} c_{jl} r_{u_{l-1}u_l}^j \leq \bar{c}_u, \forall u \in V_P\label{eqn:con-comp}\\
&\quad \sum_{j\in\mathcal{J}}\sum_{l=1}^{L^j} m_{jl} r_{u_{l-1}u_l}^j \leq \bar{m}_u, \forall u \in V_P\label{eqn:con-mem}\\
&\quad r_{uv}^j\in\{0,1\},\forall (u,v)\in E, j\in\mathcal{J},\label{eqn:con-binary-r}
\end{align}
where $\bar{c}_u$ and $\bar{m}_u$ denote the maximum amount of computation tasks that are allowed at node $u$ and the memory (such as RAM) capacity of node $u$, respectively. The objective function \eqref{eqn:obj-simdis} is the total computation plus transmission time of all the jobs. The constraint \eqref{eqn:con-path} ensures that for each job $j$, a path is found from source $s^j_0$ to destination $t^j_{L^j}$, so that the inference on the data generated at the source can be delivered to the destination. The constraints \eqref{eqn:con-comp} and \eqref{eqn:con-mem}, from \cite{baccour:RLPDNN} and \cite{disabato:DDCNN}, require that the amount of computation tasks and memory requirement at each node should not exceed certain thresholds. 

This is an equivalent reformulation of the one\footnote{There are some other constraints, but we only present the constraints that are essential to the discussion} in \cite{baccour:RLPDNN} and \cite{disabato:DDCNN}, except that our formulation does not fix a priori the path for data transfer between nodes. Our formulation can take into account the transmission time better, especially when link transmission rates are relatively low. More importantly, our formulation is an ILP which has advantages over IQP in \cite{baccour:RLPDNN} and \cite{disabato:DDCNN} (Although IQP can be linearized, it introduces a large number of variables and constraints). The number of variables in our formulation is $J\cdot |E_P|\cdot (L+1)$, whereas it is $J\cdot |V_P|\cdot L$ in the formulation of \cite{baccour:RLPDNN} and \cite{disabato:DDCNN}. Although our formulation has slightly more variables, the linearity of formulation enables to find a solution more quickly. This is demonstrated through numerical evaluation in Section \ref{sec:numerical-evaluation}.

%\begin{itemize}
%\item numerical advantages
%\end{itemize}

\subsection{Formulation for Single Job with Waiting Time}
Since the above formulation takes into account only the service time, most of the computation tasks may be assigned to fast path with high transmission capacity or node with high computation capacity. For example, if the transmission time is negligible because link capacities are considerably high compared to computation capacity, then the service time can be minimized by assigning all the jobs to the node with highest computation capacity. However, in practice, such an assignment can incur large waiting times, and hence, adversely affect the completion time. Although introducing the budget constraints \eqref{eqn:con-comp} and \eqref{eqn:con-mem} has the effect of minimizing the completion by distributing the workload over the network, one may consider the completion time more directly without such constraints.

\begin{figure}[htbp]
\centerline{\includegraphics[scale=0.55]{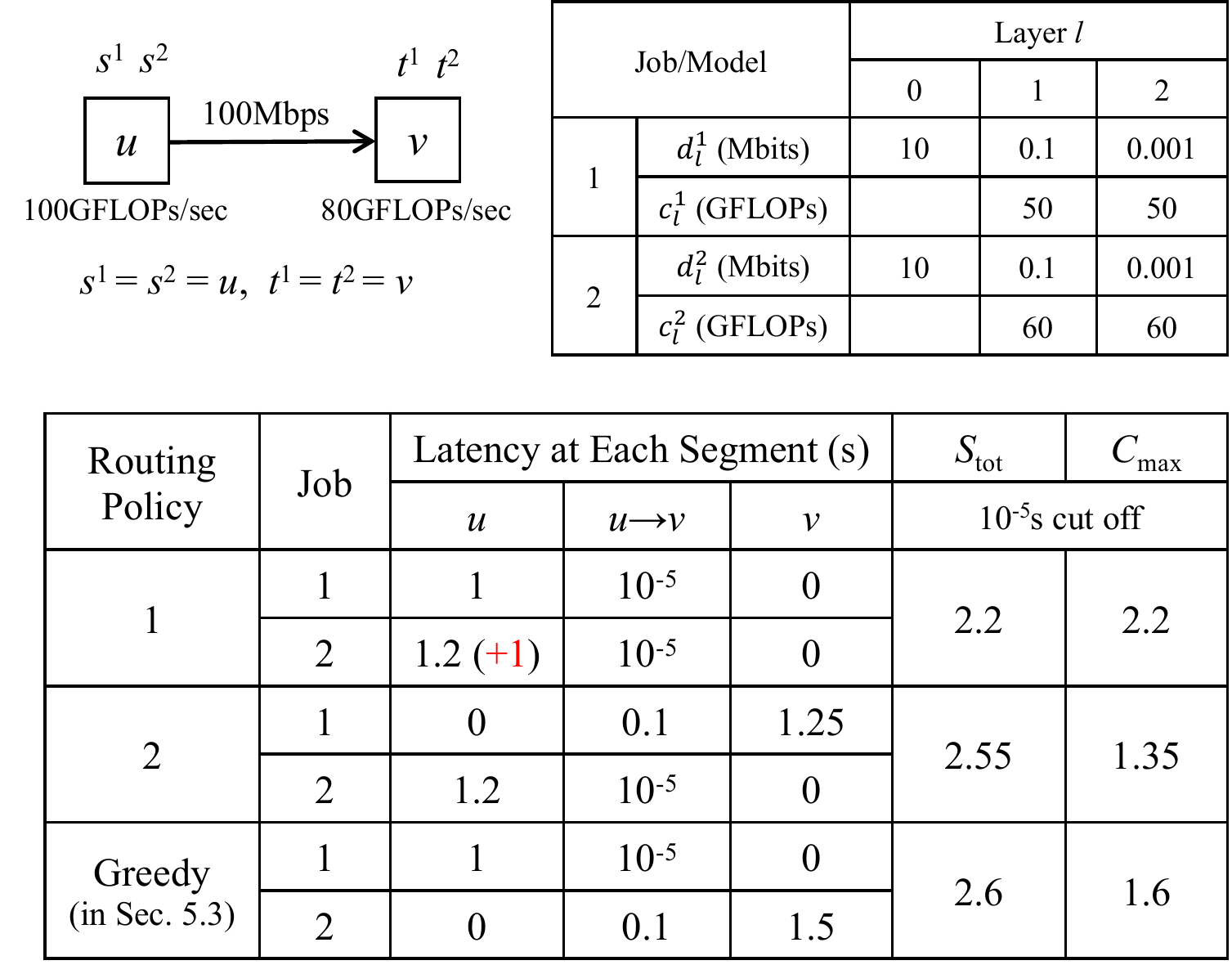}}
\vspace{-0.2cm}
\caption{Examples of routing policy. The number ``+1'' in red color indicates waiting time. $S_{\rm tot}$: total service time}
\label{fig:motivating-example-new}
\vspace{-0.1cm}
\end{figure}

For example, consider the scenario in Fig.~\ref{fig:motivating-example-new} where two models, each with two layers need to be routed. Recall that $d^j_0$ is the input data size. Routing policy 1 seeks to minimize the total service time, thereby processing all the layers at node $u$. If job 1 is processed first, then job 2 must wait until job 1 is finished, which incurs one second of waiting. This results in the completion time of 2.2s. On the other hand, routing policy 2 aims at minimizing the completion time by assigning job 1 to node $v$ and job 2 to node $u$. The total service time under this policy is larger than that under policy 1. However, the completion time is reduced to 1.35s as neither job experiences waiting.

It is important to note that the waiting time needs to be considered in order to minimize the completion time. Furthermore, if one can somehow take into account the waiting time in a more direct manner, some or all of the budget constraints (which are needed to distribute the workload) may be removed. This is because the solution that seeks to minimize waiting time will necessarily distribute the workload over the network. This is also an important aspect of considering waiting time (in a more direct manner) because it is easier to solve the problem with fewer constraints. In the following, we discuss how our framework can be applied in this context.

%In this sense, the constraints such as \eqref{eqn:con-comp} and \eqref{eqn:con-mem} are necessary so that computation tasks are not loaded only on a small set of resources. In other words, if one can somehow incorporate the waiting time in the objective function, the budget constraints \eqref{eqn:con-comp} and \eqref{eqn:con-mem} may not be necessary because they were needed for reducing waiting time. 

We start with a simple case with $J=1$, i.e., a single job needs to be routed. Recall that the queue lengths $Q_u,\forall u\in V_P$ and $Q_{uv},\forall (u,v)\in E_P$ are the computation and transmission tasks that are waiting to be computed and transmitted, respectively. These queue lengths in the physical network are reflected into the layered graph as follows:
\begin{enumerate}
\item $Q_{u_l v_l} = Q_{uv},\forall (u,v)\in E_P, l\in \mathcal{L}$
\item $Q_{u_{l-1} u_l} = Q_u, \forall u\in V_P, l=1,...,L$
\end{enumerate}
For brevity, let us omit the superscript $j$. Consider the following formulation:

{\small
\begin{align}
\min_{r,z} &\quad \sum_{(u,v)\in E} \frac{q_{uv}}{\mu_{uv}} r_{uv} + \sum_{(u,v)\in E\setminus E_C} \frac{Q_{uv}}{\mu_{uv}} r_{uv} + \sum_{u\in V_P}  \frac{Q_u}{\mu_u} z_u\label{eqn:objective-single}\\
\mbox{s.t.} & \quad z_u\geq r_{u_{l-1}u_l},\forall u\in V_P, l=1,...,L\label{eqn:constraint-node-selection}\\
&\quad \sum_{v:(u,v)\in E} r_{uv} -  \sum_{v:(v,u)\in E} r_{vu} = \begin{cases}
1,& \mbox{if } u = s_0\\
-1,& \mbox{if } u = t_L\\
0,& \mbox{otherwise}
\end{cases}, \forall u\in V\label{eqn:constraint-flow-conservation}\\
&\quad r_{uv}\in\{0,1\},\forall (u,v)\in E \label{eqn:constraint-binary-r}\\
&\quad z_{u}\in\{0,1\},\forall u\in V_P \label{eqn:constraint-binary-z}
\end{align}
}

Similar to \eqref{eqn:con-path}, the constraint \eqref{eqn:constraint-flow-conservation} ensures that the solution finds a path from source to destination, including the node at which each layer of model is to be computed. The first term in \eqref{eqn:objective-single} is the total service time. The second and third terms represent the waiting time at links and nodes, respectively, provided that upon the arrival at corresponding components, $Q_{uv}$ and $Q_u$ are the actual amount of data waiting for transmission/computation. Specifically, the second term adds the waiting times at all the links traversed. Note that if the solution translates to a path with cycle(s) in the original graph $G_P$, then the second term may add $Q_{uv}$ multiple times for some $(u,v)\in E_P$. By constraint \eqref{eqn:constraint-node-selection},  $z_u$ is 1 in the optimal solution only if node $u\in V_P$ is selected for computing layer(s). The third term is thus the total waiting time at nodes. Likewise, if the solution gives a loopy path, then there may be a node visited twice or more, and hence, the third term may not give the actual waiting time. Otherwise, if the optimal solution gives a simple path, then the objective function value is the actual end-to-end latency. This is summarized below.
%Note that it is possible for the optimal solution to give a path with cycle(s) in the original graph. In this case, the second term can potentially add the waiting time at a link twice or more, and the third term may add

\begin{lemma}\label{lem:simple-path-latency}
Assume that the optimal solution $r=[r_{uv},(u,v)\in E]$ of problem \eqref{eqn:objective-single}-\eqref{eqn:constraint-binary-z} translates to a simple path (i.e., a path with no cycles) in the original graph. Then, the objective function value in \eqref{eqn:objective-single} is the job completion time.
\end{lemma}
In simulations, we were able to observe that a simple path is found in many of the instances tested.

The problem \eqref{eqn:objective-single}-\eqref{eqn:constraint-binary-z} is an ILP which is NP-hard in general. However, we show that our formulation has a special structure called total unimodularity, so that the optimal solution to ILP can still be found with the relaxation of binarity constraints in \eqref{eqn:constraint-binary-r} and \eqref{eqn:constraint-binary-z} as $0\leq r_{uv}\leq 1$ and $0\leq z_u\leq 1$. With this relaxation, the formulation is just a linear program (LP) which is polynomial time solvable. In general, LP introduces fractional solution, which in our problem, implies multiple paths with fractional flow. For example, the solution can have the form of $r_{uv}=0.7$ along one path and $r_{uv}=0.3$ along another, while the goal is to find a single path with unit flow. The following theorem shows that our formulation always has an integral optimal solution.
\begin{theorem}\label{thm:LP-relaxation-works}
The LP relaxation of formulation \eqref{eqn:objective-single}-\eqref{eqn:constraint-binary-z} has an integral optimal solution (i.e., an optimal single path with minimum completion time).
\begin{IEEEproof}
See Appendix \ref{app:proof-theorem-LP-relaxation-works}.
\end{IEEEproof}
\end{theorem}
By this theorem, one can find a solution of the ILP \eqref{eqn:objective-single}-\eqref{eqn:constraint-binary-z} by solving its LP relaxation which is much easier. This can be used to develop an efficient algorithm for routing multiple jobs, as presented in the following.

%This theorem can be used to develop an efficient algorithm for routing multiple jobs, as presented in the following.

{\bf Remark.} The quantities $Q_u$ and $Q_{uv}$ can be replaced with expected tasks when the job arrives at the corresponding resource, so that expected waiting time can be taken into account. The computation requirement $q_{u_{l-1}u_l}=c_l$ of layer $l$ can be replaced with $c_l(1+\alpha \mathbb{I}_{\{m_l>M_u\}})$ where $M_u$ is the physical memory capacity of node $u$, and $\mathbb{I}$ is the indicator function taking the value 1 if the condition is satisfied, and 0 otherwise. This reflects the slowdown of computation if the memory requirement exceeds the physical memory and techniques such as swapping are activated to accommodate excessive memory demand. This way, one can take into account the impact of memory demand, without budget constraints such as \eqref{eqn:con-mem}, which is likely to make the problem easier to solve.

\subsection{Greedy Algorithm for Priority-based Job Routing}
In order to show how the formulation \eqref{eqn:objective-single}-\eqref{eqn:constraint-binary-z} can be used, we assume preemptive scheduling at links and nodes. Priority is assigned to every inference ``job'' (not to individual layers), and uniformly applied to all the layers belonging to the same job. For instance, consider two jobs $j_1$ and $j_2$, and suppose that job $j_1$ has higher priority than job $j_2$. The computation/transmission task of a layer in job $j_2$ can be preempted (while it is being computed/transmitted) up on the arrival of the computation/transmission task of job $j_1$. 

We consider a greedy algorithm that determines the routing and priority of jobs, one at a time in the order of priority. Suppose now that the queue lengths $Q_u,\forall u\in V_P$ and $Q_{uv},\forall (u,v)\in E_P$ represent the computation and transmission tasks (that are already routed) with higher priority than the current model to be routed. For routing a job, we use the formulation \eqref{eqn:objective-single}-\eqref{eqn:constraint-binary-z}. Again, the first term in \eqref{eqn:objective-single} represents the total service time at nodes and links. The third term is the total waiting time at nodes, assuming that the computation of the current job at each node $u$ must wait until all the computation tasks of higher priority. This is obviously an upper bound on the actual waiting time because some of the computation tasks in $Q_u$ may have been finished by the time when the computation task of the current job arrives at the node. Similarly, the second term in \eqref{eqn:objective-single} is an upper bound on the total waiting time at links. The objective function in \eqref{eqn:objective-single} is thus an upper bound on the actual service time plus waiting time which is equal to the completion time. Therefore, in this case, the formulation \eqref{eqn:objective-single}-\eqref{eqn:constraint-binary-z} seeks to find a routing that minimizes an upper bound on the completion time of current job.

Note that due to networking delays (i.e., transmission times), it is hard to express the actual waiting time in a simple form. The upper-bound approach enables a simple formulation. Although there may be a gap between upper bound and actual completion time, we expect that minimizing the upper bound would have the effect of minimizing the actual job completion time. In the following, we consider this fictitious system in which the upper bound is treated as the actual waiting time.

Let $Q=[Q_{uv},\forall (u,v)\in E]$ which is the vector of unfinished transmission/computation tasks in the network. Recall that this vector determines the waiting time of jobs. Let $C^j(Q)$ be the optimal objective function value of formulation \eqref{eqn:objective-single}-\eqref{eqn:constraint-binary-z} for job $j$. That is, $C^j(Q)$ is the completion time of job $j$ if it is routed based on the formulation in the presence of unfinished tasks $Q$. Let $r^*(j)$ be the optimal routing variable in this case. Note that by Theorem \ref{thm:LP-relaxation-works}, $C^j(Q)$ and its solution $r^*(j)$ can be found by solving the LP relaxation of \eqref{eqn:objective-single}-\eqref{eqn:constraint-binary-z}.

Algorithm \ref{alg:greedy} shows the greedy policy. First, it computes the completion time of every job, and selects the job with earliest completion time (line 1). This selected job is given highest priority. Second, the unfinished task vector $Q$ is updated (line 2) so that the remaining jobs with lower priority can be routed with updated waiting time. The same procedure is repeated until all the jobs are routed. As shown in Fig.~\ref{fig:motivating-example-new}, the greedy algorithm finds a solution that balances the workload over the network, and achieves the completion time close to the minimum possible value which is achieved by routing policy 2. Although the algorithm is derived assuming the fictitious system, we expect that the greedy algorithm performs well in practice as it penalizes the solution utilizing the resources that are highly occupied.

%\SetKwProg{Init}{init}{}{}
\SetKwInput{KwInit}{Init}
\SetKwInput{KwGiven}{Given}
\SetKwInput{KwOutput}{Output}

\begin{algorithm}%[H]
\SetAlgoLined
\KwGiven{Jobs $\mathcal{J}=\{1,...,J\}$
}
\KwInit{$Q_{uv}=0,\forall (u,v)\in E$; $U = \mathcal{J}$; $p=1;$}
\While{$U \neq \emptyset$}{
$j_p = \arg\min\limits_{j\in U} C^j(Q)$\;
$Q_{uv} \leftarrow Q_{uv} + q_{uv}^{j_p} r_{uv}^*(j_p),\forall (u,v)\in E$\;
$p \leftarrow p+1$\;
$U \leftarrow U\setminus \{j_p\}$\;
}
\KwOutput{ Priority \& Routing: $[j_1>\cdots>j_J]$ \& $[r^*(j_p), \forall p=1,...,J]$\
%\KwOutput{ Priority: $j_1>\cdots>j_J$\\
%\hspace{1.45cm}Routing: $r_{uv}^*(j_p), \forall (u,v), p=1,...,J$\
}
\caption{Greedy Algorithm}\label{alg:greedy}
\end{algorithm}

The approximation ratio of this algorithm can be  derived. Let us abuse the notation by defining $|V_P|$ and $|E_P|$ as the numbers of nodes with positive computation capacity and edges with finite transmission capacity respectively in the original graph $G_P$. Let $T^*$ be the minimum possible job completion time in the actual system.
\begin{theorem}\label{thm:greedy-analysis}
Assume that the graph $G_P$ is $k$-edge-connected and the formulation \eqref{eqn:objective-single}-\eqref{eqn:constraint-binary-z} always finds a simple path in $G_P$. Then, the job completion time under Algorithm \ref{alg:greedy} in the actual system is at most $\alpha T^*$, where
{\small
\begin{align*}
&\alpha = \max\left\{2\alpha_{tx}, \frac{2(L+1)(|V_P|+|E_P|)\alpha_{tx}}{k}\right.,\\
&\qquad\qquad\qquad \left.\left(1 + \frac{|E_P|}{|V_P|}\right)\alpha_{cp}\right\}\left\{2-\frac{1}{|V_P|+|E_P|}\right\}\\
&\alpha_{tx} = \frac{h_L\cdot\max\limits_{j,l} d_l^j \cdot \max\limits_{(u,v)\in E_P}\,\, \mu_{uv}}{h_S\cdot\min\limits_{j,l} d_l^j \cdot \min\limits_{(u,v)\in E_P} \,\,\mu_{uv}}, \,\,
\alpha_{cp} = \frac{\max_{u\in V_P}\mu_u}{\min_{u\in V_P}\mu_u}\\
&h_L = \max_j h_L^j, \,\, h_S = \min_j h_S^j
%&h_L^j = \mbox{longest path length in $G_P$ in hop count btw. $s^j$ and $t^j$}
\end{align*}
%where
\begin{align*}
%&\alpha_{tx} = \frac{h_L\cdot\max\limits_{j,l} d_l^j \cdot \max\limits_{(u,v)\in E_P}\,\, \mu_{uv}}{h_S\cdot\min\limits_{j,l} d_l^j \cdot \min\limits_{(u,v)\in E_P} \,\,\mu_{uv}}, \,\,
%\alpha_{cp} = \frac{\max_{u\in V_P}\mu_u}{\min_{u\in V_P}\mu_u}\\
%&h_L = \max_j h_L^j, \,\, h_S = \min_j h_S^j\\
&h_L^j = \mbox{longest path length in $G_P$ in hop count btw. $s^j$ and $t^j$}\\
&h_S^j = \mbox{shortest path length in $G_P$ in hop count btw. $s^j$ and $t^j$.}
\end{align*}}
\begin{IEEEproof}
See Appendix \ref{app:proof-theorem-greedy-analysis}.
\end{IEEEproof}
\end{theorem}
Therefore, the greedy policy is an $\alpha$-approximation algorithm. 

\begin{corollary}\label{cor:two-approx}
With zero network delay (or infinite link capacity) and identical computation capacity at all nodes, the greedy policy is a $(2-1/|V_P|)$-approximation algorithm.

\begin{IEEEproof}
See Appendix \ref{app:proof-corollary-2-approx}.
\end{IEEEproof}
\end{corollary}

Although the approximation ratio in \ref{thm:greedy-analysis} seems loose, in the special case as in Corollary \ref{cor:two-approx}, our algorithm approximates the optimal solution within a factor smaller than 2. We thus anticipate our algorithm performs well in many other scenarios. We show in Section \ref{sec:numerical-evaluation} that in many cases, the greedy approximates the optimal in the fictitious system within a small neighborhood.

% {\color{red} Recall that the completion time under the greedy policy is an upper bound on the completion time that will be achieved by greedy in the actual system. Hence, the approximation ratio of greedy immediately leads to the approximation of greedy in the actual system. We thus anticipate our algorithm performs well in many other scenarios.}

{\bf Remark.} The priority-based scheduling discussed in this section may not be realistic in general communication networks, however such a scheduling policy will likely to gain increasing attention. As the demand for supporting a large spectrum of applications increases,  the conventional one-size-fit-all network solution becomes no longer viable. The concept of network slicing has always been one of the key service paradigms that can solve this problem. With network slicing, certain services are assigned logically isolated computing and networking resource blocks, within which customized scheduling policy can be employed. The primary goal of network slicing is to provide a certain level of quality-of-service (QoS) guarantee through flexible scheduling in the slice. For this reason, priority-based scheduling and admission control (into slice) is an important problem of research \cite{wang:energy,jiang:network}. Ubiquitous AI service provider may purchase a network slice, and get subscribers with differentiated service level agreement. Accordingly, we envision that the discussion of priority-based scheduling for inference can potentially be an important issue in practice in the future.

\section{Numerical Evaluation}\label{sec:numerical-evaluation}
In this section, we evaluate numerical advantages of our framework and the performance of the greedy policy in Algorithm \ref{alg:greedy}. We adopt the settings from \cite{disabato:DDCNN} where IoT environment is considered. The original physical network $G_P$ is generated by random geometric graph. That is, nodes are randomly placed over the plane of $x \times x$ square with $x=30$m. For a pair of nodes, there is a bidirectional link if the two nodes are within certain distance, e.g., communication range of each other in wireless settings. Such a communication range is set to $7.5$m. If a disconnected graph is generated, it is discarded to consider only connected graphs. The link transmission rate $\mu_{uv}$ between two neighbors is randomly set to one of the values in $\{1,...,5\}\times \frac{\gamma\cdot\mu_{tx}^{\rm max}}{5}$ where $\gamma$ is the scaling parameter and $\mu_{tx}^{\rm max}$ is set to 72.2Mbps which is the data rate of WiFi 4. We scale up ($\gamma\uparrow$) and down ($\gamma\downarrow$) the entire link transmission rates in order to examine our framework in various regimes where the computation time dominates the completion time over transmission time or the other way around. %Table~\ref{tbl:parameters} shows the values of these parameters.

%\begin{table}[!t]
%% increase table row spacing, adjust to taste
%%\renewcommand{\arraystretch}{1.3}
%% if using array.sty, it might be a good idea to tweak the value of
%% \extrarowheight as needed to properly center the text within the cells
%\caption{Parameters}
%\label{tbl:parameters}
%\centering
%\begin{tabular}{|c||c|}
%\hline
%Name & Value\\
%\hline\hline
%$\mu_{tx}^{\rm max}$ (maximum transmission rate) & 72.2Mbps\\
%\hline
%$x$ (length of one side of plane) & 30m\\
%\hline
%communication range & 7.5m\\
%\hline
%\end{tabular}
%\end{table}

We assume three types of IoT nodes including Orange Pi Zero (OPZ), Beaglebone AI (BAI) and Raspberry Pi 3 (RP3). Table~\ref{tbl:node-types} shows the specs of these devices. Note that $\bar{c}_u$ is just the parameter introduced by the constraint \eqref{eqn:con-comp}, not a real hardware spec. Each node in $G_P$ is randomly set to one of the three types.%, which determines the computation capacity $\mu_u$ of the node. 

\begin{table}[!t]
\caption{Node types and capacities. MM: million multiplications}
\label{tbl:node-types}
\centering
\begin{tabular}{c||c|c|c}
\hline
Node type & $\bar{c}_u$ (MM) & $\bar{m}_u$ (MB) & $\mu_u$ (MM/s)\\
\hline\hline
OPZ & 100,000 & 524.288 & 360\\
\hline
BAI & 100,000 & 131.072 & 480\\
\hline
RP3 & 100,000 & 524.288 & 560\\
\hline
\end{tabular}
\end{table}

We consider three CNN models including SevenLayerNet (SLN), AlexNet (AN) and ResNet101 (RN). The memory $m^j_l$ and computation $c^j_l$ requirement and output data size $d^j_l$ of these models are identical to those in \cite{disabato:DDCNN} where compressed form of layers is assumed (see Table~\ref{tbl:cnn-models}). For each job, its source and destination are chosen randomly from $G_P$, and one of the three models is randomly selected.

\begin{table*}[!t]
\caption{CNN models and parameters. MM: million multiplications, KB: KiloBytes}
\label{tbl:cnn-models}
\centering
\begin{tabular}{c|c||c|c|c|c|c|c|c|c|c|c|c}
\hline
\multicolumn{2}{c||}{Model} & \multicolumn{10}{c|}{Requirement of each layer $l=0,...,L^j$ ($d^j_0$: input data size)}& Unit\\
\hline\hline
\multirow{3}{*}{SLN} & $d^j_l$ &9.41 & 50.18 & 12.54 & 1.54 & 0.77 & 0.04 & - & - & - & - & KB\\
\cline{2-13}
& $c^j_l$ & - & 3.81 & 20.08 & 1.20 & 0.07 & 0.002 & - & - & - & - & MM\\
\cline{2-13}
& $m^j_l$ & - & 19.20 & 409.60 & 4816.90 & 294.91 & 7.68 & - & - & - & - & KB\\
\hline
\multirow{3}{*}{AN} & $d^j_l$ & 618.35 & 279.94 & 173.06 & 259.58 & 259.58 & 36.86 & 16.38 & 16.38 & 4.00 & - & KB\\
\cline{2-13}
& $c^j_l$ & - &105.73 & 224.34 & 149.52 & 112.14 & 74.84 & 37.75 & 16.78 & 4.10 & - & MM\\
\cline{2-13}
& $m^j_l$ & - &139.78 & 1229.82 & 3540.48 & 2655.74 & 1770.50 & 151011.39 &  67125.25 & 16388.00 & - & KB\\
\hline
\multirow{3}{*}{RN} & $d^j_l$ & 602.12 & 802.82 & 802.82 & 200.71 & 50.18 & 50.18 & 50.18 & 50.18 & 12.54 & 4.00 & KB\\
\cline{2-13}
& $c^j_l$ & - & 118.01 & 616.56 & 757.86 & 950.53 & 1156.06 & 1156.06 & 1156.06 & 565.18 & 2.20 & MM\\
\cline{2-13}
& $m^j_l$ & - & 37.63 & 786.43 & 2228.22 & 21757.95 & 26738.69 & 26738.69 & 26738.69 & 51380.22 & 8388.61 & KB\\
\hline
\end{tabular}
\end{table*}

\subsection{Example of Solution Found by Our Framework}
\begin{figure}[htbp]
\captionsetup[subfloat]{justification=centering}
\centering
\subfloat[Original graph with source-destination pairs]{
\includegraphics[scale=0.45]{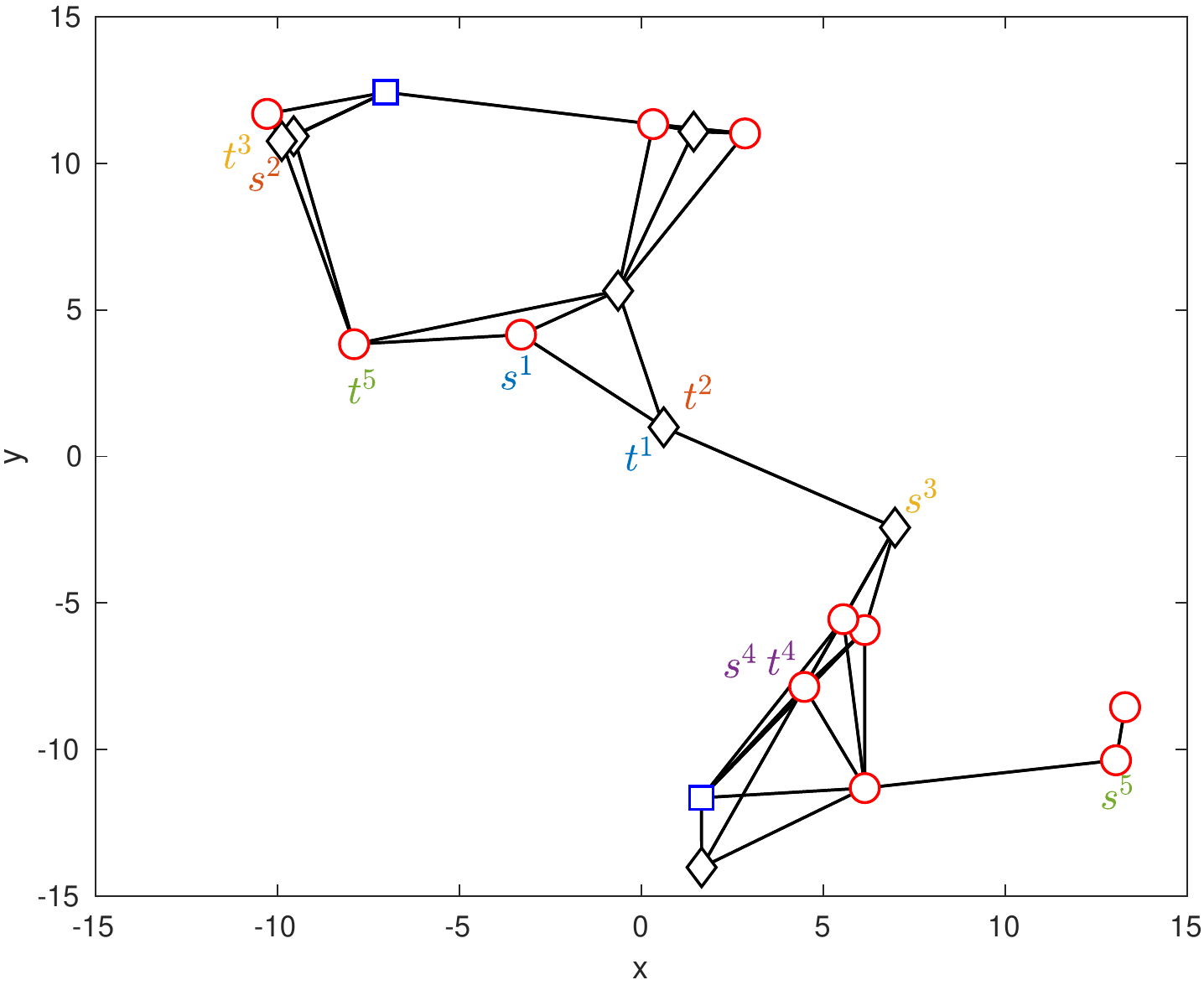}
\label{fig:solution-example-original-graph}
}
\hfill
\subfloat[Layered graph with source-destination pairs]{
\includegraphics[scale=0.45]{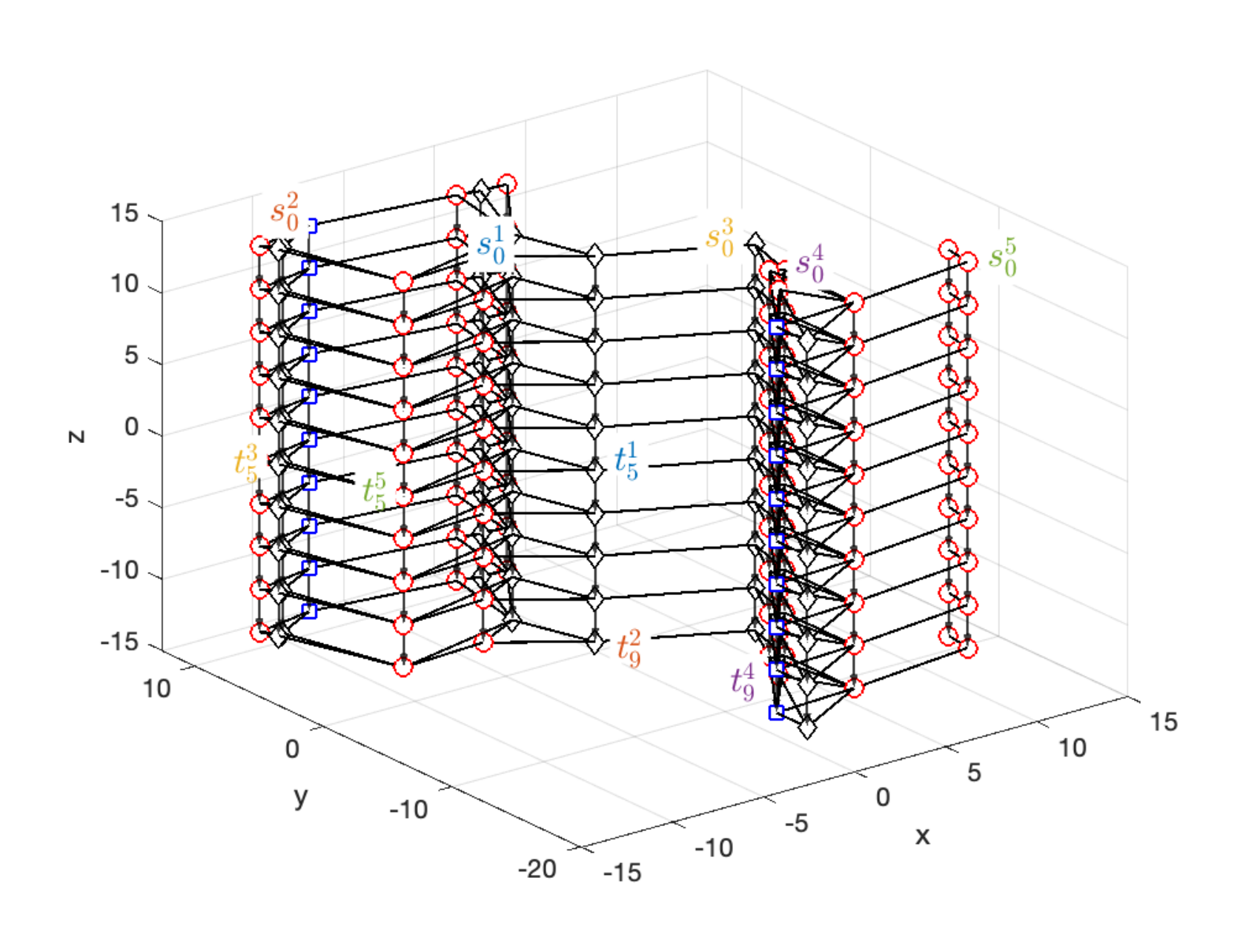}
\label{fig:solution-example-layered-graph}
}
\hfill
\subfloat[Paths found in layered graph]{
\includegraphics[scale=0.45]{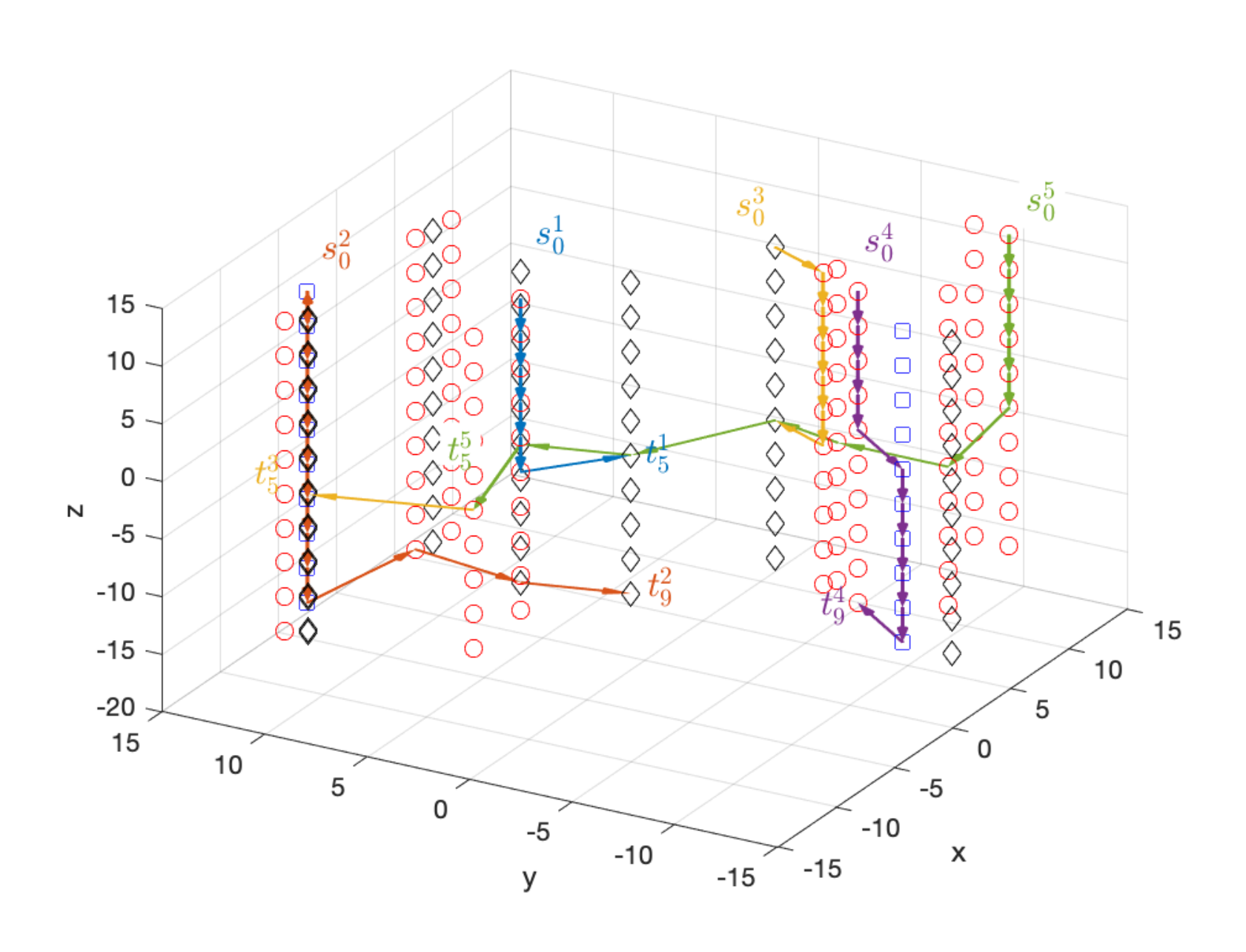}
\label{fig:solution-example-layered-path}
}
\hfill
\subfloat[Paths mapped to original graph]{
\includegraphics[scale=0.45]{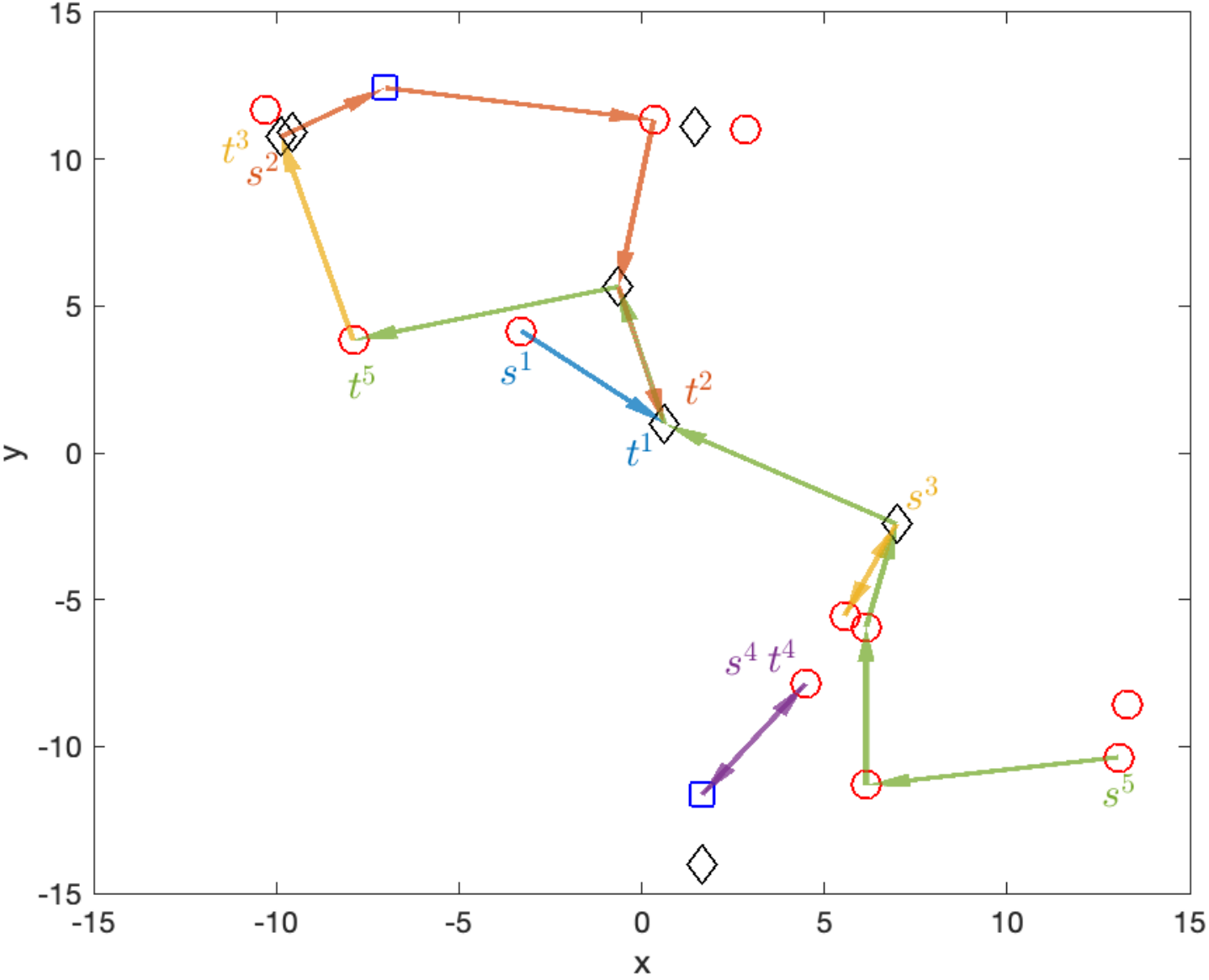}
\label{fig:solution-example-original-path}
}
\vspace{-0.1cm}
\caption{Example of how the solution is found. Formulation \eqref{eqn:obj-simdis}-\eqref{eqn:con-binary-r} is used to solve a problem with $|V_P|=20, J=5$.}
\label{fig:solution-example-procedure}
\end{figure}

We first present a small example of how our framework finds a solution for distributing DNN inference jobs. There are 20 nodes and 5 jobs. Fig. \ref{fig:solution-example-original-graph} shows the original network graph with the source-destination pair of each job to be routed. Different markers indicate different node types (circle: OPZ, diamond: BAI, square: RP3). Note that the source and destination of job 4 are the same node, implying that the corresponding node generates the data and needs the inference result on the data.

Next, the layered graph is constructed as shown in Fig.~\ref{fig:solution-example-layered-graph}. Note that the 3D geometry is only for visualizing the graph, and not needed in the process of finding a solution. The sources are placed in layer 0, i.e., $G_0$, while the destinations are placed in the final layer of corresponding DNN model. For instance, job 1 has 5 layers in the model, and hence, its destination is placed at $t^1_5$ in layer 5.  By solving the formulation, the path is found for each source-destination pair in the layered graph.

Fig.~\ref{fig:solution-example-layered-path} depicts the five paths found. The straight down arrow from upper layer to lower layer indicates that the corresponding layer is computed at the corresponding node (in the original graph). Note that job 4 is processed at the the source node up to layer 4, and then at the neighboring square node up to the final layer. The result is then delivered back to the source (which is also the destination). Every other job is processed at a single node. These paths in the layered graph are then mapped back to the original graph as shown in Fig.~\ref{fig:solution-example-original-path}.

%\begin{figure*}[!t]
%\centering
%\subfloat[]{\includegraphics[width=2.5in]{fig1}%
%\label{fig_first_case}}
%\hfil
%\subfloat[]{\includegraphics[width=2.5in]{fig1}%
%\label{fig_second_case}}
%\caption{Dae. Ad quatur autat ut porepel itemoles dolor autem fuga. Bus quia con nessunti as remo di quatus non perum que nimus. (a) Case I. (b) Case II.}
%\label{fig_sim}
%\end{figure*}

%\begin{figure}[htbp]
%\captionsetup[subfloat]{justification=centering}
%\centering
%\begin{subfloat}[b]{0.45\textwidth}
%\centering
%\includegraphics[scale=0.45]{figures/path-example-N=20_J=5_original_graph}
%\vspace{-0.0cm}
%\caption{Original graph with source-destination pairs}
%\label{fig:solution-example-original-graph}
%\end{subfloat}\vspace{0.3cm}
%\begin{subfloat}[b]{0.45\textwidth}
%\centering
%\includegraphics[scale=0.45]{figures/path-example-N=20_J=5_layered_graph}
%\vspace{-0.1cm}
%\caption{Layered graph with source-destination pairs}
%\label{fig:solution-example-layered-graph}
%\end{subfloat}\vspace{0.3cm}
%\begin{subfloat}[b]{0.45\textwidth}
%\centering
%\includegraphics[scale=0.45]{figures/path-example-N=20_J=5_layered_path2}
%\vspace{-0.1cm}
%\caption{Paths found in layered graph}
%\label{fig:solution-example-layered-path}
%\end{subfloat}\vspace{0.3cm}
%\begin{subfloat}[b]{0.45\textwidth}
%\centering
%\includegraphics[scale=0.45]{figures/path-example-N=20_J=5_original_path}
%\vspace{-0.1cm}
%\caption{Paths mapped to original graph}
%\label{fig:solution-example-original-path}
%\end{subfloat}
%\vspace{-0.1cm}
%\caption{Example of how the solution is found. Formulation \eqref{eqn:obj-simdis}-\eqref{eqn:con-binary-r} is used to solve a problem with $|V_P|=20, J=5$.}
%\label{fig:solution-example-procedure}
%\end{figure}

\subsection{Numerical Advantages of Our Framework}
We compare our ILP formulation \eqref{eqn:obj-simdis}-\eqref{eqn:con-binary-r} with the IQP formulation in \cite{disabato:DDCNN}. The two formulations solve the same problems. We use Gurobi Optimizer running on Intel Xeon Gold 5220R CPU @ 2.2GHz with 24 cores and 256GB of RAM.  Recall that in the formulation of \cite{disabato:DDCNN}, the transmission latency between nodes is assumed to be shortest hop distance $*$ (data size $\div$ constant rate). Since links can have different rates, the constant rate in the IQP formulation is set to the average rate of links on the shortest-hop path. The number of nodes $|V_P|$ is fixed to 50, while the number of jobs $J$ and the data rate scaling $\gamma$ are varied as $J\in\{5, 10, 20, 30, 40\}$ and $\gamma\in\{0.2, 0.5, 1.0, 2.0\}$.

\begin{figure}[htbp]
%\captionsetup[subfloat]{justification=centering}
\centering
\subfloat[Relative gap of total service time (objective function value in \eqref{eqn:obj-simdis}) defined as $\frac{O_{\rm IQP} - O_{\rm ILP}}{O_{\rm ILP}}\times 100$(\%)]{
\includegraphics[scale=0.4]{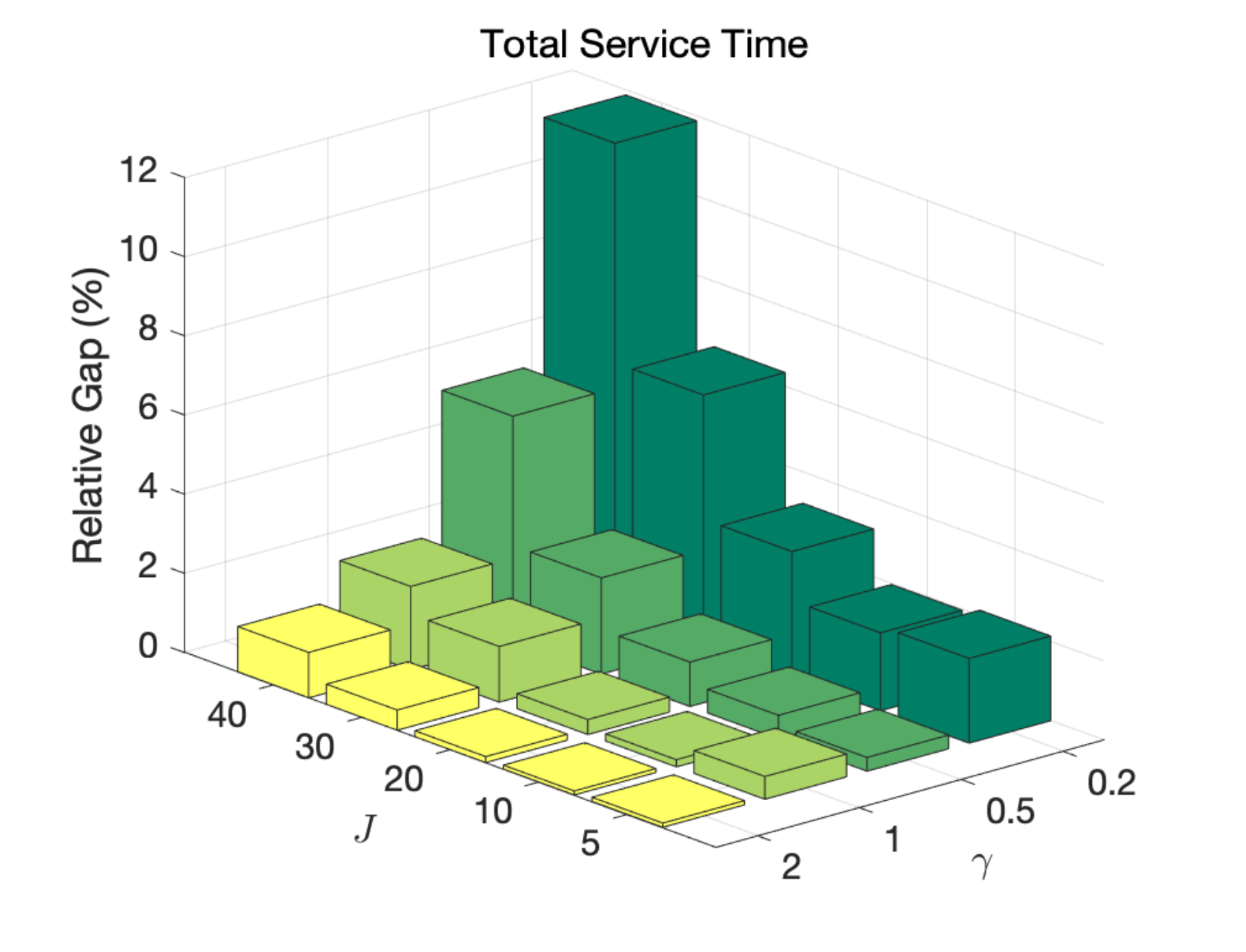}
\label{fig:comparison-relative-gap}
}
\hfill
\subfloat[Runtime to solve formulation]{
\includegraphics[scale=0.4]{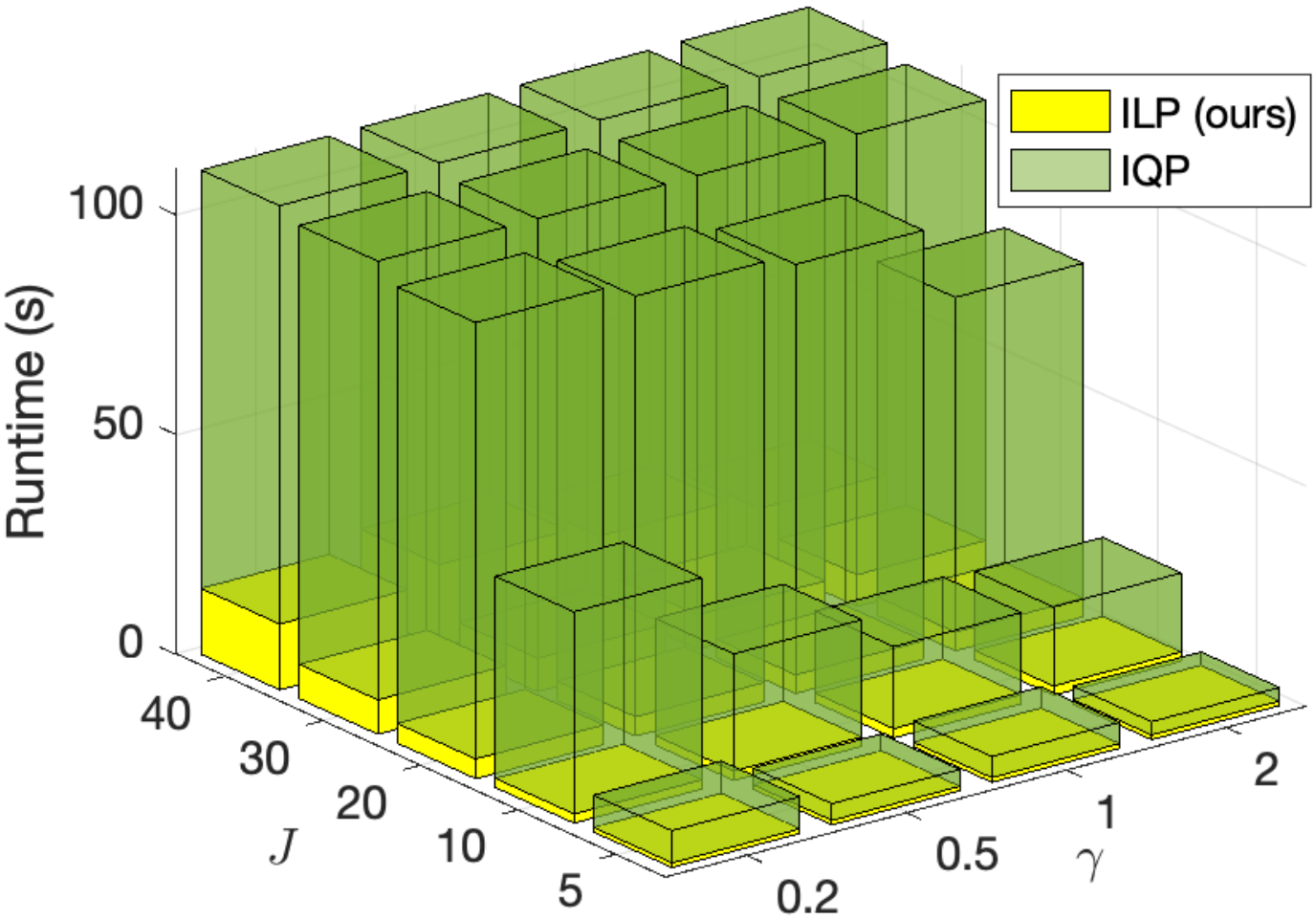}
\label{fig:comparison-runtime}
}
\vspace{-0.1cm}
\caption{Comparison of ILP (ours) and IQP, with $|V_P|=50$. Each point is the average of the results from 50 random problem instances}
\label{fig:comparison-with-IQP}
\end{figure}

Fig.~\ref{fig:comparison-with-IQP} compares the solutions found by our ILP formulation and IQP formulation of \cite{disabato:DDCNN}. In Fig.~\ref{fig:comparison-relative-gap}, the service times (i.e., the objective function value in \eqref{eqn:obj-simdis}) are shown in the form of relative gap defined as $\frac{O_{\rm IQP} - O_{\rm ILP}}{O_{\rm ILP}}\times 100$(\%) where $O_{\rm F}$ is the optimal objective function value of formulation F. For small values of $\gamma$, link transmission rates are small, and hence, the transmission latency takes a significant portion of service time. Since our formulation takes into account both transmission and computation latency, the solution found by our formulation achieves smaller service time especially when $\gamma$ is small. On the other hand, for large values of $\gamma$, the transmission latency becomes less significant, and the performance gap between ours and IQP decreases. Observe also that the performance gap increases as the number of jobs $J$ increases. The gap in each job's service time cumulates more with a large number of jobs.

The IQP formulation can be easily fixed to better take into account transmission time and hence, the service time of IQP formulation can be improved. We believe what is more important is the runtime to solve formulation. Since it takes nearly indefinite time for some problem instances, the time limit of the solver is set to 100s for both formulations. Fig. \ref{fig:comparison-runtime} shows the runtime of both formulations. As the number of jobs increases, the IQP quickly hits the time limit, meaning that in 100 seconds, it fails to find a solution which can be claimed to be optimal. On the other hand, our formulation finds an optimal solution much faster than IQP, with slowly growing runtime in the number of jobs.

\subsection{Integrality Gap: Comparison with LP Relaxation}
In order to further investigate the numerical advantages of our framework, we evaluate the integrality gap. Consider the LP relaxation of formulation \eqref{eqn:obj-simdis}-\eqref{eqn:con-binary-r}, i.e., $r^j_{uv}\in\{0,1\}$ relaxed as $0\leq r_{uv}^j\leq 1$. Let $O_{\rm LP}$ be the optimal objective function value of LP relaxation. Clearly, we have $O_{\rm LP}\leq O_{\rm ILP}$ because LP relaxation optimizes over the superset of what is considered in ILP. For minimization problem, the integrality gap is defined as $\frac{O_{\rm ILP}}{O_{\rm LP}}$, which is always no smaller than 1. If the solver finds an integral solution which yields objective function value equal to $O_{\rm LP}$, it means that it has found an optimal solution. However, this happens only when the integrality gap is 1. Therefore, the integrality gap is a crucial property of formulation from numerical perspectives (as well as approximation algorithms).

\begin{figure}[htbp]
%\captionsetup[subfloat]{justification=centering}
\centering
\subfloat[$\gamma=0.2$]{
\includegraphics[scale=0.3]{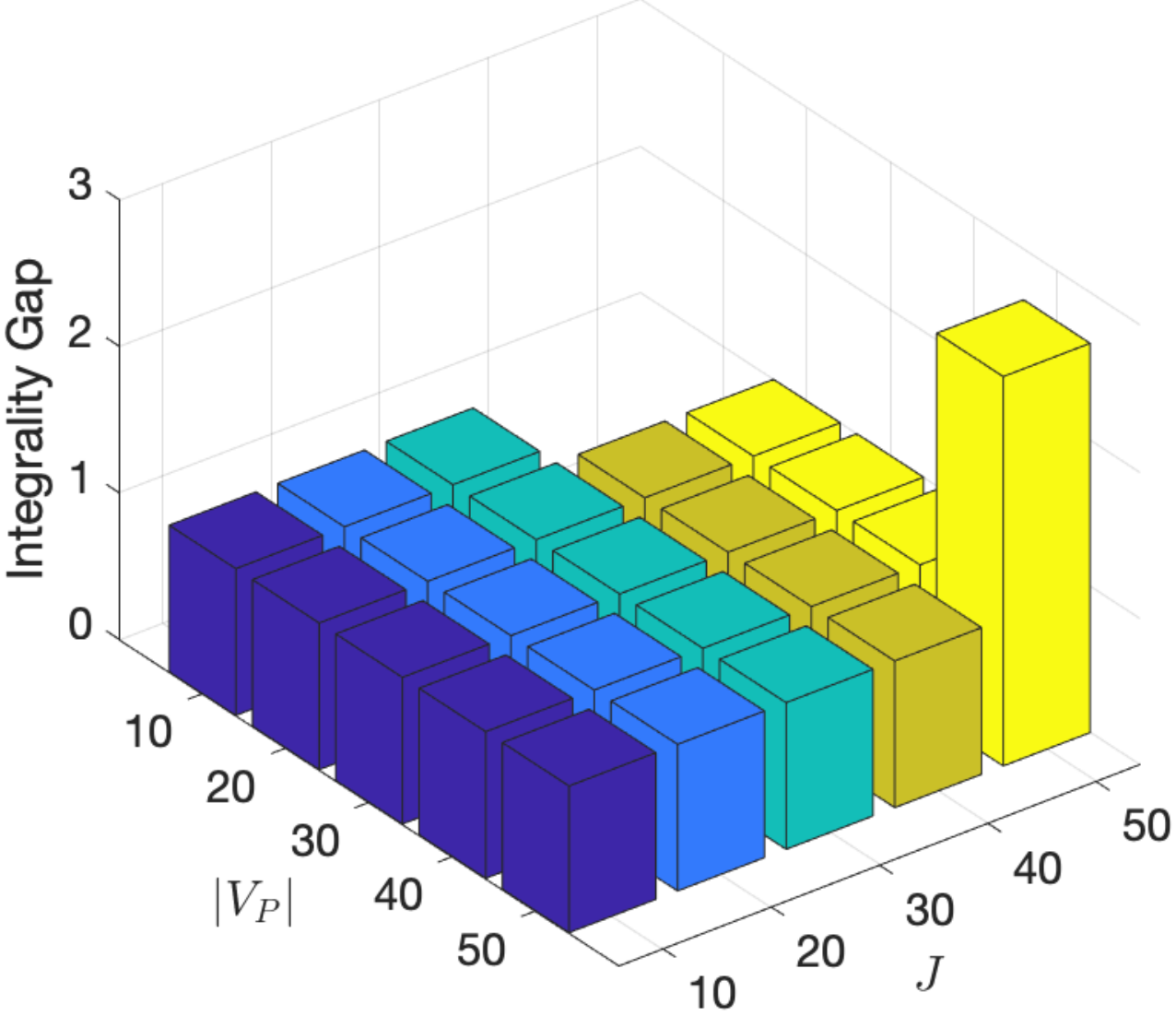}
\label{fig:integrality-gap-gamma=0.2}
}
\subfloat[$\gamma=2$]{
\includegraphics[scale=0.3]{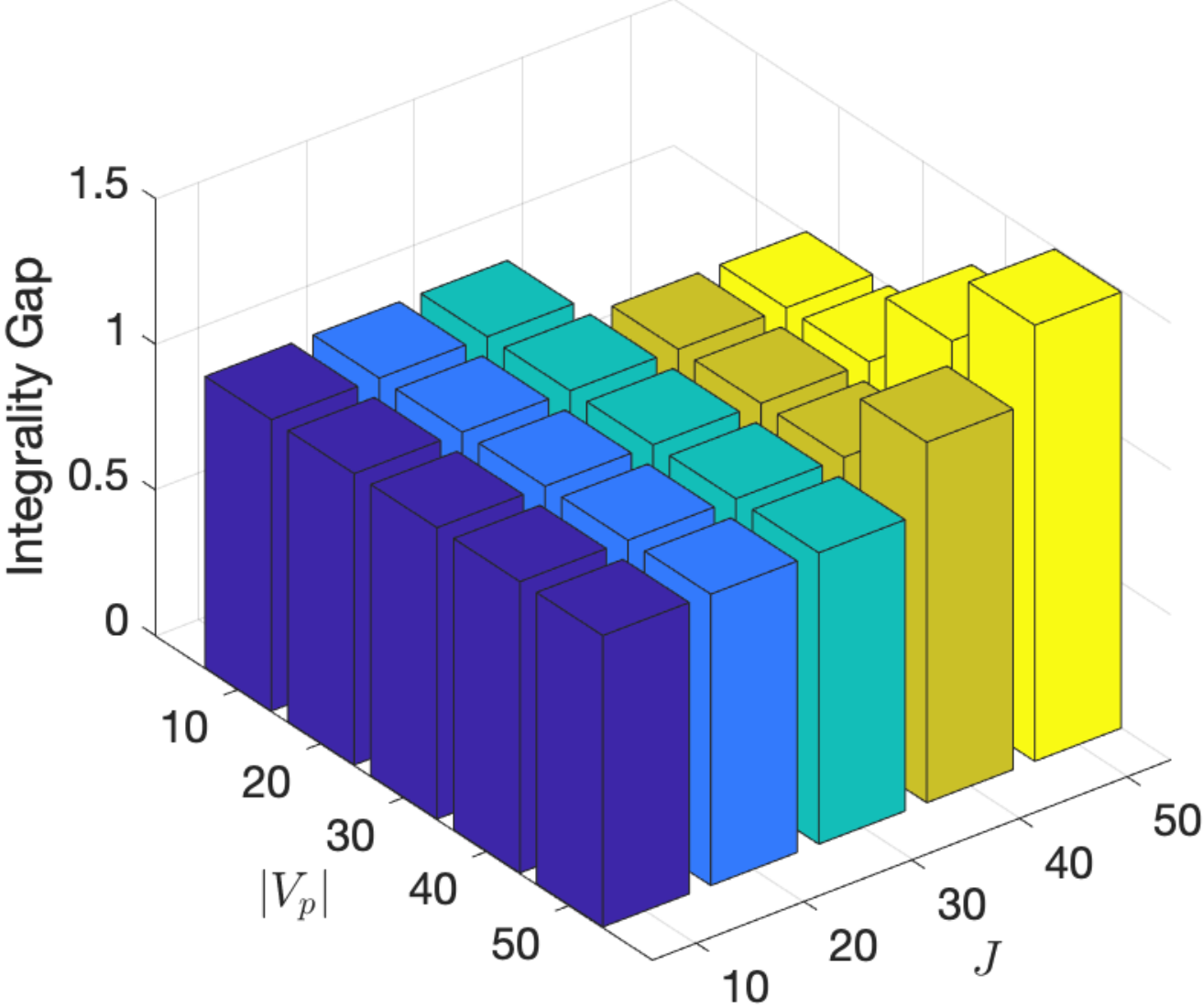}
\label{fig:integrality-gap-gamma=2}
}
\vspace{-0.1cm}
\caption{Integrality gap of our ILP formulation \eqref{eqn:obj-simdis}-\eqref{eqn:con-binary-r}. Each point is the average of the results from 50 random problem instances. The results with other values of $\gamma$ are omitted due to limited space.}
\label{fig:integrality-gap}
\end{figure}

Fig.~\ref{fig:integrality-gap} shows the integrality gap under various circumstances\footnote{In contrast with the above evaluation, some problem instances turn out to be infeasible especially when the network with small size (i.e., small $|V_P|$) needs to serve a large number of jobs. We omit the results for the combination of $(|V_P|,J)$ with which there are fewer than 10 feasible instances out of 50}. For most of the problem instances, the integrality gap is nearly 1, and we postulate that the gap mostly results from numerical precision issue as the objective function values of ILP and LP are often equal upto a few digits after the decimal point. As mentioned above, the integrality gap of 1 can accelerate the solver to find an optimal solution. The results in Fig.~\ref{fig:integrality-gap} explain why our formulation is able to find an optimal solution much faster than IQP.

Note that for some instances, the integrality gap is relatively large. For example, with $(J=50,|V_P|=50)$ in Fig.~\ref{fig:integrality-gap-gamma=0.2},  the average integrality gap is about 3. We found that for only two instances out of 50, the ILP finds an extraordinarily long path with cycles in the layered graph, after the time limit 100s of the solver is reached. The objective function values in the two cases are about 9724s and 7725s respectively, while the LP relaxation attains about 169s and 181s respectively. Such a solution of loopy path can never be optimal because one can immediately remove cycle(s) in the path to obtain a solution with smaller objective function value and reduced burden in the budget constraints. Although we believe this is an issue of solver, such an extraordinary solution can be avoided by adding a small penalty such as $\delta \sum_{j,(u,v)} r^j_{uv}$ with small positive $\delta$. This penalizes the solution with cycle(s), forcing the solver to find an acyclic path.

\begin{figure}[htbp]
\centering
\includegraphics[scale=0.4]{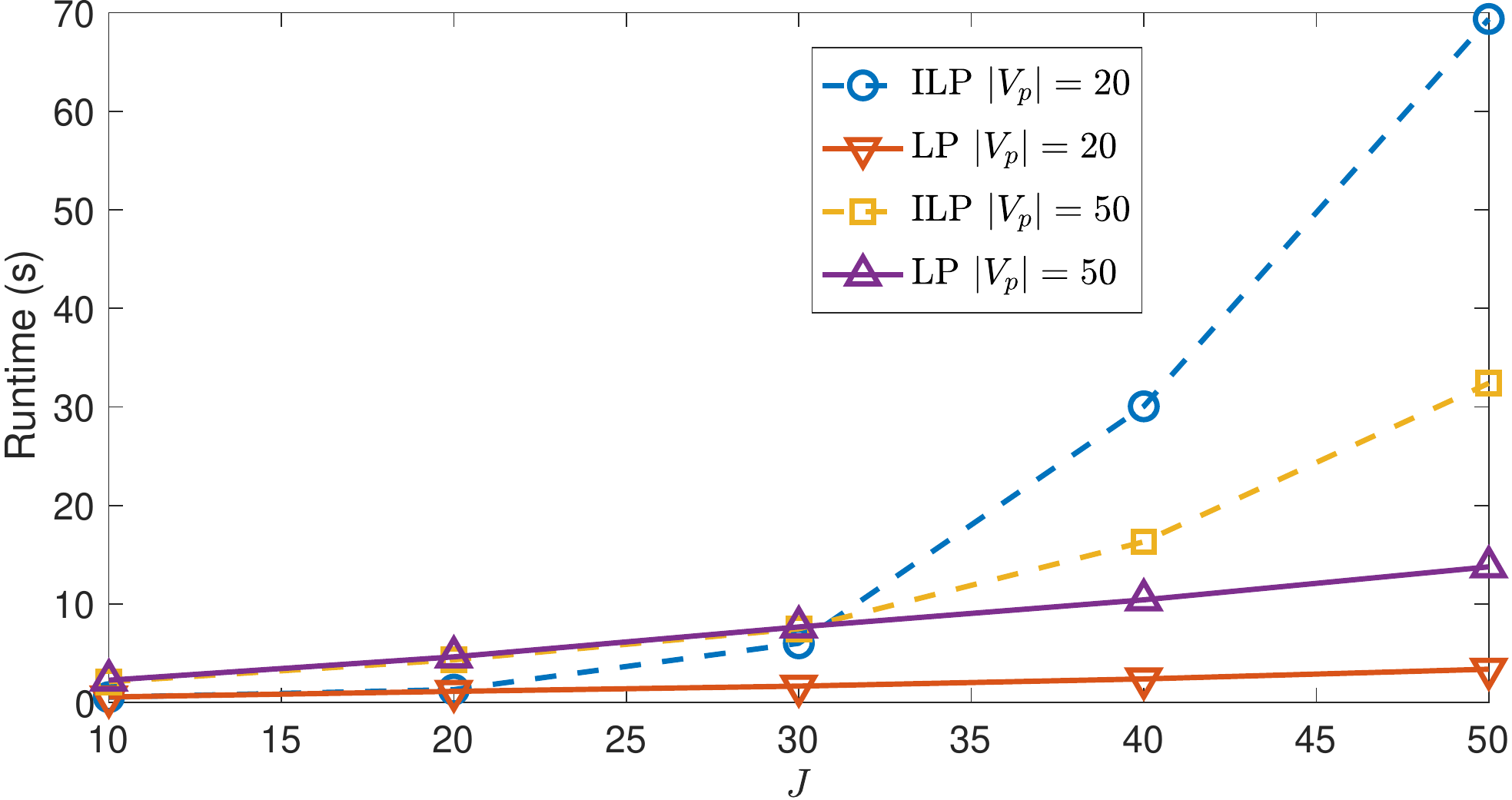}
\caption{Runtime of ILP and LP vs. number of jobs $J$}
\label{fig:integrality-gap-runtime}
\end{figure}

\begin{figure}[htbp]
%\captionsetup[subfloat]{justification=centering}
\centering
\subfloat[$|V_P|=20, J=40$: only 5 jobs' paths are shown.]{
\includegraphics[scale=0.45]{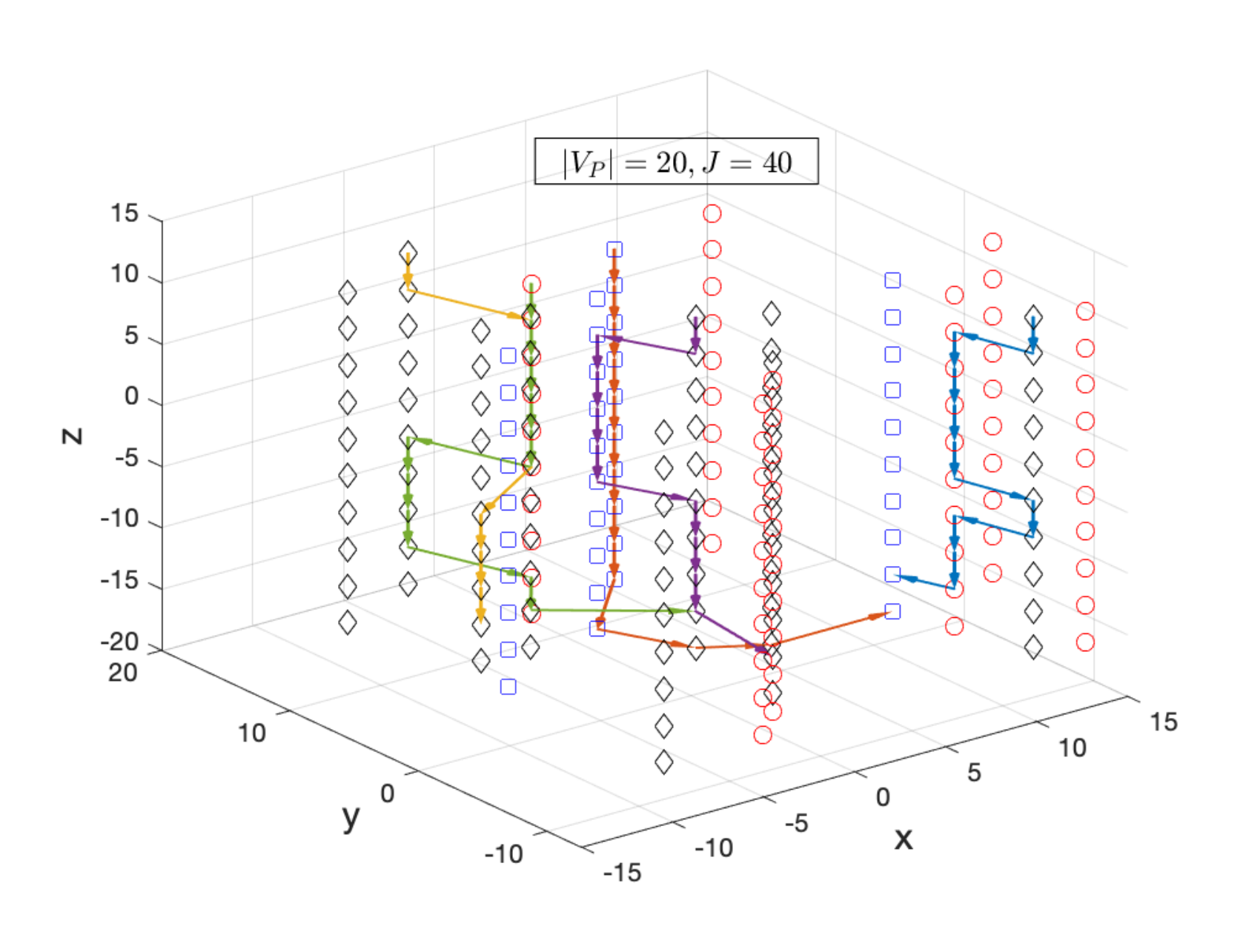}
%\label{fig:integrality-gap-gamma=0.2}
}
\hfill
\subfloat[$|V_P|=20, J=40$: only 5 jobs' paths are shown. Nodes are omitted to avoid congested figure.]{
\includegraphics[scale=0.45]{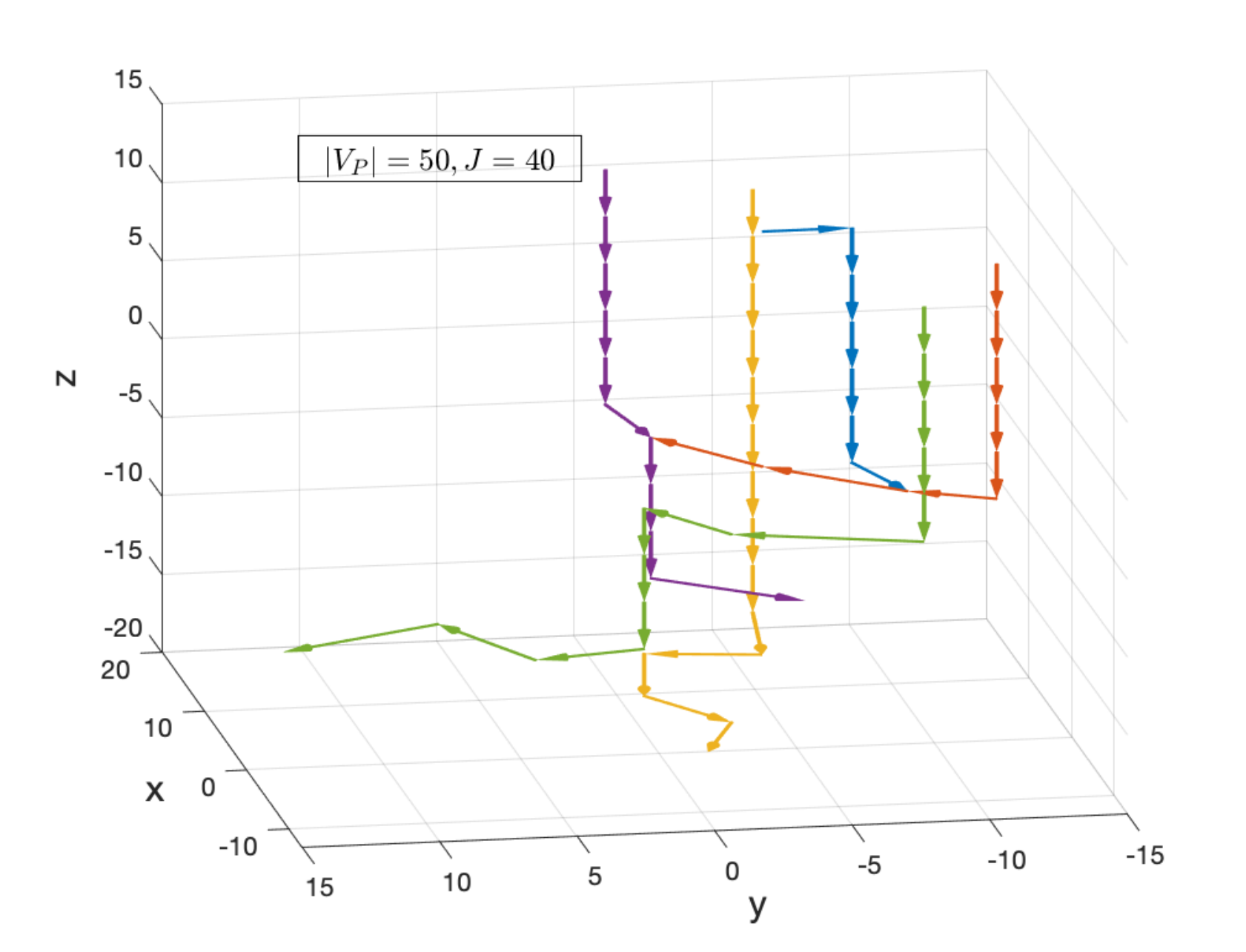}
%\label{fig:integrality-gap-gamma=2}
}
\vspace{-0.1cm}
\caption{Integrality gap of our ILP formulation \eqref{eqn:obj-simdis}-\eqref{eqn:con-binary-r}. Each point is the average of the results from 50 random problem instances. The results with other values of $\gamma$ are omitted due to limited space.}
\label{fig:path-example-large-small}
\end{figure}

Fig.~\ref{fig:integrality-gap-runtime} compares the runtime. As the number of jobs increase, the runtime of ILP increases relatively fast compared to that of LP. Notice that it takes more time for ILP to find a solution when the number of jobs is large and the network size is small. For instance, the runtime with $(|V_P|=50,J=40)$ is smaller than that with $(|V_P|=20,J=40)$. This is because when the resources are abundant in the large network, it is easier to find a solution to support all the jobs. On the other hand, if the resources are scarce compared to the demand, the solution tends to be complicated, which obviously takes the solver  a long time to discover such a solution. Fig.~\ref{fig:path-example-large-small} compares the solutions under small and large values of $|V_P|$ with fixed $J$, where only the first (in job-index-wise) 5 jobs' paths are shown. When $|V_P|$ is small,  the paths tend to be complicated, and indeed, many of the paths are cyclic in the original graph. It is easy to guess from the form of the solution that it tries to pack the paths to stay below the budgets. On the other hand, when $|V_P|$ is large, all the paths are simple in the original graph. This example clearly shows why it may take longer to solve the problem when $|V_P|$ is small and $J$ is large.

%Gap of IQP meaningless; relaxation is nonconvex QP, and has worse obj val than IQP. This doesn't give useful info about the formulation

\subsection{Evaluation of Greedy Algorithm}
We now evaluate the greedy policy in Algorithm \ref{alg:greedy}. The following two algorithms are considered for comparison with greedy, denoted by GRD.
\begin{enumerate}
\item Optimal algorithm (denoted by OPT) that finds a path for each job such that the completion time is minimized in the fictitious system. The formulation is given in Appendix \ref{app:opt-algo}.
\item Node-first selection algorithm (denoted by NFS) that selects a single node, say $u$, for each job with earliest computation completion, and takes the shortest (in transmission completion time) path from source to $u$ and from $u$ to destination. The details are given in Appendix \ref{app:nfs-algo}.
\end{enumerate}
While OPT finds an optimal solution, it may take an excessively long time. In contrast, NFS may find a solution quickly, but at the expense of increased completion time. 

\begin{figure}[htbp]
%\captionsetup[subfloat]{justification=centering}
\centering
\subfloat[Relative gap of completion time defined as $\frac{C_{\rm GRD} - C_{\rm OPT}}{C_{\rm OPT}}\times 100$(\%)]{
\includegraphics[scale=0.45]{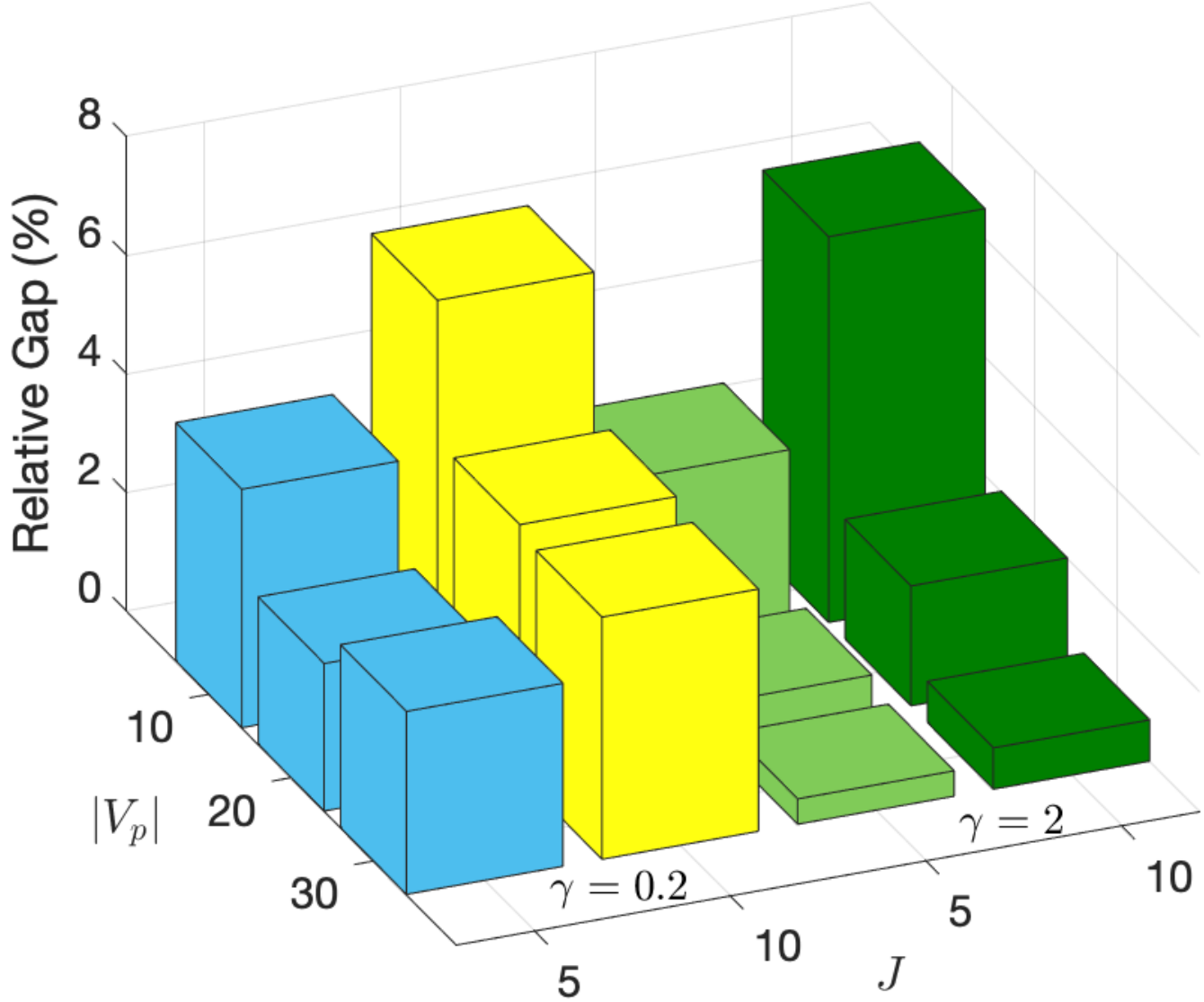}
\label{fig:evaluate-greedy-vs-opt-ctime}
}
\hfill
\subfloat[Runtime]{
\includegraphics[scale=0.45]{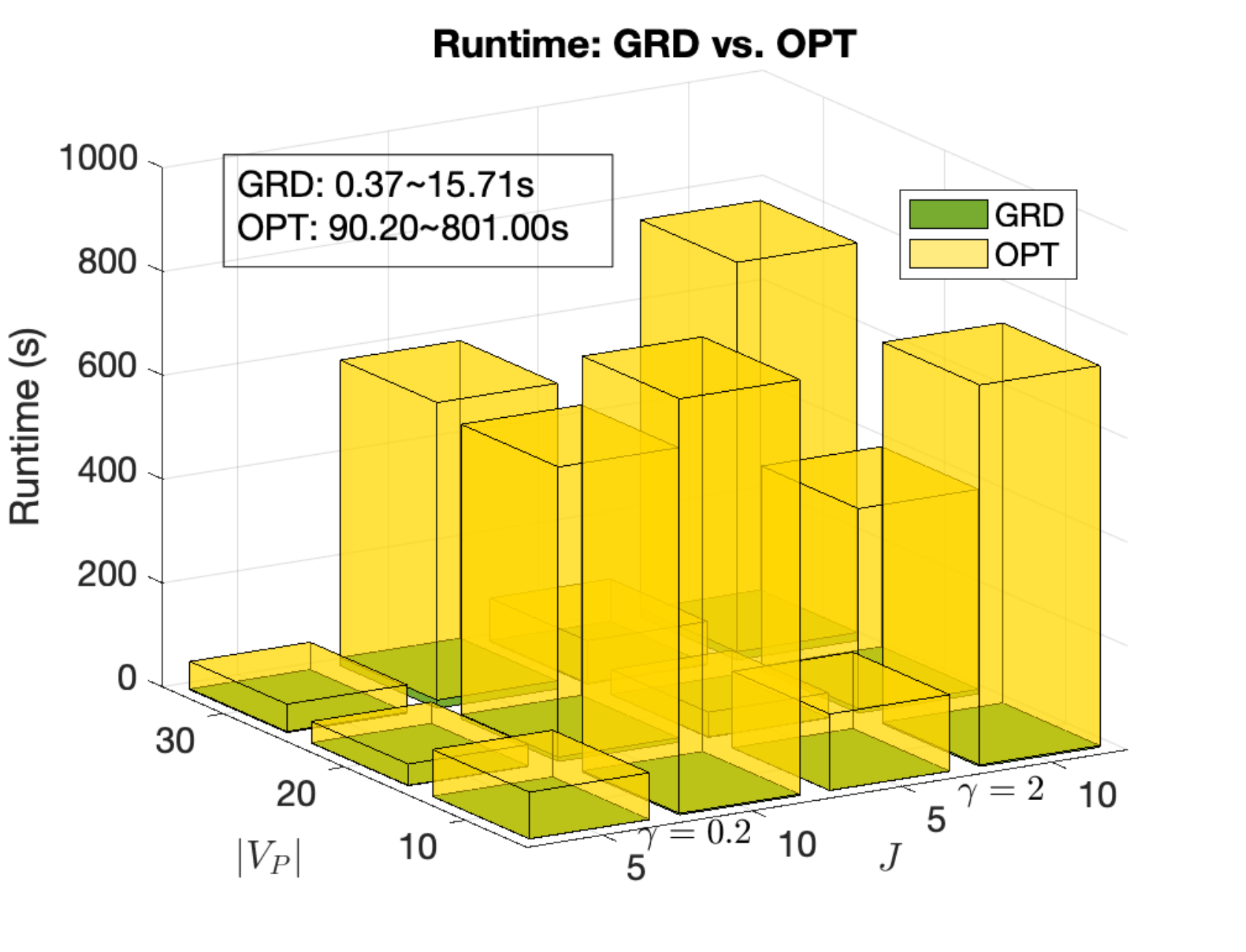}
\label{fig:evaluate-greedy-vs-opt-runtime}
}
\vspace{-0.1cm}
\caption{Comparison of greedy and optimal algorithms}
\label{fig:evaluate-greedy-vs-opt}
\end{figure}

We first compare greedy and optimal algorithms. Let $C_{\rm ALG}$ be the completion time of algorithm ALG in the fictitious system. We found that OPT fails to find a solution in 30 minutes if either $J>10$ or $N>30$. In addition, even if it finds a solution after the time limit of 30 minutes, it achieves even larger completion time than GRD or NFS. Hence, the results from those instances were dropped. In other words, we only consider the results in which the completion time of OPT is no greater than that of GRD and NFS. Fig.~\ref{fig:evaluate-greedy-vs-opt} shows those results. In Fig. \ref{fig:evaluate-greedy-vs-opt-ctime} the relative gap of completion time, defined as $\frac{C_{\rm GRD} - C_{\rm OPT}}{C_{\rm OPT}}\times 100$(\%), remains below 7\%, implying that on average, GRD achieves completion time at most 1.07 times that of OPT. As mentioned above, OPT introduces a large number of variables and constraints, and thus, it takes a long time to find an optimal solution (see Fig. \ref{fig:evaluate-greedy-vs-opt-runtime}). Note that when OPT hits the time limit of 30 minutes, most of the solutions (if found) give meaningless completion time. As mentioned above, those cases were omitted, and thus, the runtime of OPT shown in Fig. \ref{fig:evaluate-greedy-vs-opt-runtime} somewhat underestimates the true runtime. This result shows that the greedy policy is able to find nearly optimal solutions in reasonable time (at least for the cases shown in the figure).

\begin{figure}[htbp]
%\captionsetup[subfloat]{justification=centering}
\centering
\subfloat[Relative gap of completion time defined as $\frac{C_{\rm NFS} - C_{\rm GRD}}{C_{\rm GRD}}\times 100$(\%)]{
\includegraphics[scale=0.45]{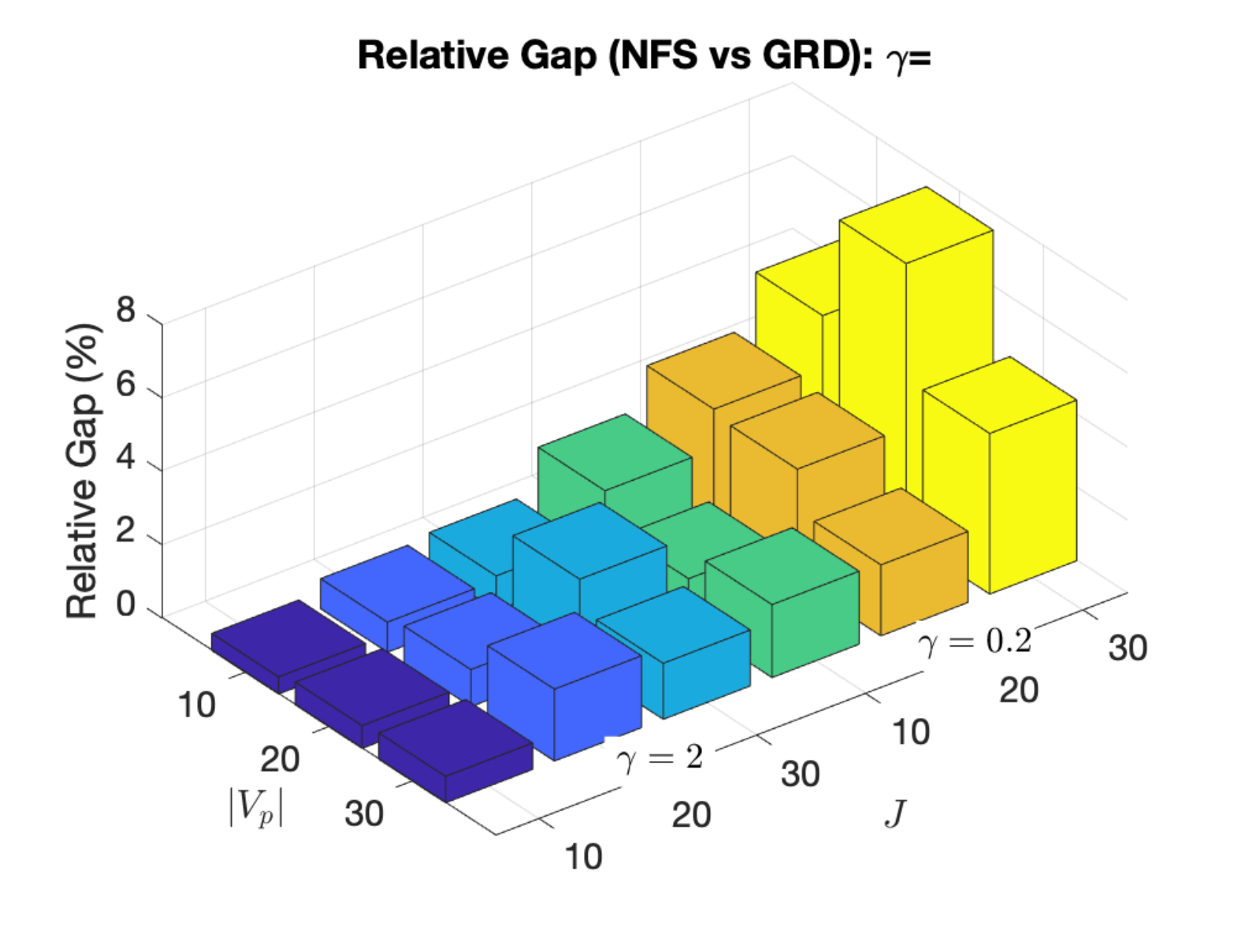}
\label{fig:evaluate-greedy-vs-nfs-ctime}
}
\hfill
\subfloat[Runtime]{
\includegraphics[scale=0.45]{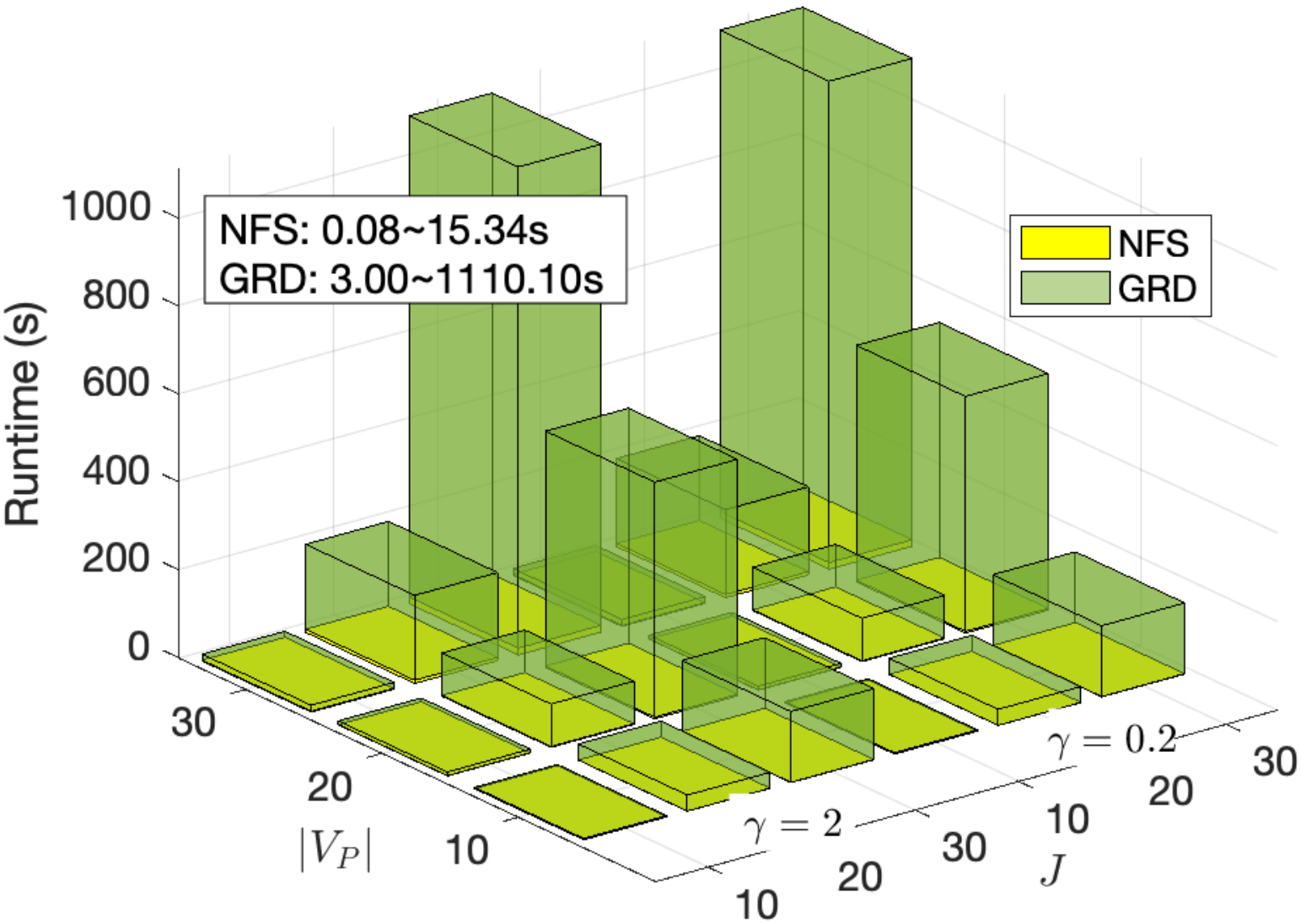}
\label{fig:evaluate-greedy-vs-nfs-runtime}
}
\vspace{-0.1cm}
\caption{Comparison of greedy and node-first selection algorithms}
\label{fig:evaluate-greedy-vs-nfs}
\end{figure}

Next, we compare greedy and node-first selection algorithms. In Fig.~\ref{fig:evaluate-greedy-vs-nfs-ctime}, the relative gap of completion time increases as the data transmission rate (small $\gamma$) decreases. This is because NFS selects a node (and thus a path) with an emphasis on computation time, while with small $\gamma$, the transmission time becomes a substantial element affecting completion time. The greedy policy finds a path adaptively to the circumstances regarding computation and transmission, thereby achieving better completion time performance. 

On the other hand, as the numbers of nodes and jobs increase, the runtime of greedy scales poorly compared to the node-first selection algorithm. The greedy requires to solve the LP relaxation of \eqref{eqn:objective-single}-\eqref{eqn:constraint-binary-z} $O(J^2)$ times. We found that as the iteration continues, it tends to take longer to solve the LP. With $J=30$, even one second of average solver runtime for single LP can result in several hundreds of seconds of total runtime. Therefore, it may be worthwhile to delve into the algorithmic solution for \eqref{eqn:objective-single}-\eqref{eqn:constraint-binary-z}, which we leave as future study.

\section{Conclusion}\label{sec:conclusion}
In this paper, we proposed a framework for routing DNN inference jobs over the distributed computing network. Our framework uses a novel layered graph model to simplify and integrate the communication and computation problems via conventional routing. Our algorithms presented in this paper show an example of what one can do by exploiting the proposed framework. Namely, a simple but effective solution for enjoying the capability of DNN even with the lack of computing power can be derived. In addition to the results in this paper, we believe that our framework can facilitate practical solutions to the problem of routing inference jobs. Our framework will therefore be able to help overcome limited computing power at end devices when ubiquitous AI is about to be in service.

%\section{Discussion??}
%\begin{itemize}
%\item additional delay due to memory shortage
%\end{itemize}

%Use the abbreviation ``Fig.~\ref{fig}'', even at the beginning of a sentence.

%\begin{table}[htbp]
%\caption{Table Type Styles}
%\begin{center}
%\begin{tabular}{|c|c|c|c|}
%\hline
%\textbf{Table}&\multicolumn{3}{|c|}{\textbf{Table Column Head}} \\
%\cline{2-4} 
%\textbf{Head} & \textbf{\textit{Table column subhead}}& \textbf{\textit{Subhead}}& \textbf{\textit{Subhead}} \\
%\hline
%copy& More table copy$^{\mathrm{a}}$& &  \\
%\hline
%\multicolumn{4}{l}{$^{\mathrm{a}}$Sample of a Table footnote.}
%\end{tabular}
%\label{tab1}
%\end{center}
%\end{table}

%\begin{comment}
\appendices
\section{Proof of Theorem \ref{thm:LP-relaxation-works}}\label{app:proof-theorem-LP-relaxation-works}
We start with some background needed to show the aforementioned property. The definitions and facts in the following can be found in \cite{nemhauser:integer}.

\begin{definition}
An $m \times n$ matrix $A$ is \emph{totally unimodular} (TU) if the determinant of each square submatrix is equal to 0, 1 or -1.
\end{definition}
\begin{definition}
A \emph{polyhedron} $P\subseteq \mathbb{R}^n$ is the set of points that satisfy a finite number of linear inequalities, that is, $P=\{x\in \mathbb{R}^n: Ax \leq b\}$ where $A\in\mathbb{R}^{m\times n}$ and $b\in\mathbb{R}^m$.
\end{definition}
\begin{definition}
A nonempty polyhedron is said to be \emph{integral} if all of its extreme points\footnote{A point in polyhedron $P$ is extreme if it cannot be expressed as a nontrivial convex combination of two distinct points in $P$.} are integral.
\end{definition}

The following two lemmas are the key to our analysis.
\begin{lemma}\label{lem:TU-to-integrality}
If $A\in\mathbb{R}^{m\times n}$ is TU, then $P(b)=\{x\in\mathbb{R}^n_+ : Ax \leq b\}$ is integral for all $b \in \mathbb{Z}^m$ for which it is nonempty (the same is true for $P(b)=\{x\in\mathbb{R}^n_+ : Ax=b\}$).
\end{lemma}
\begin{lemma}\label{lem:LP-integrality}
For a polyhedron $P\subseteq \mathbb{R}^n$, a solvable LP ($\max\{c^T x: x\in P\}$) has an integral optimal solution for all $c \in \mathbb{R}^n$ if and only if $P$ is integral.
\end{lemma}
By Lemmas \ref{lem:TU-to-integrality} and \ref{lem:LP-integrality}, if the constraint matrix is TU, then solving LP with integral vector $b$ in $P$ gives an integral optimal solution (or if not, the fractional solution can be rounded to an integral solution without losing optimality). In the case of binary ILP, LP relaxation replaces the constraints $(\cdot \in \{0,1\})$ with $(0\leq \cdot \leq 1)$. If the polyhedron with this relaxation is integral, then solving the LP relaxation gives a binary optimal solution because all the variables are constrained to be between 0 and 1 and hence integrality implies binarity. We exploit this fact to show that LP relaxation of our formulation has a binary optimal solution which in our case is a single path.

\subsection{Matrix Representation of Our Formulation}
We first represent the formulation in a matrix form. Let $y=[z; r_1; r_2]\footnote{The notation $;$ is used to indicate the beginning of a new row. For example, $[1; -1; 1]$ is a 3-dimensional column vector.}\in\mathbb{R}^{(L+1)(|V_P|+|E_P|)}$ where $z = [z_u,\forall u\in V_P] \in \mathbb{R}^{|V_P|}$, $r_1=[r_{u_{l-1}u_l},\forall u\in V_P, l=1,2,...,L]\in \mathbb{R}^{L\cdot|V_P|}$ and $r_2=[r_{uv},\forall (u,v)\in E_l, l=0,1,2,...,L]\in \mathbb{R}^{(L+1)\cdot|E_P|}$. Let $A_1$ and $A_2$ be the matrices corresponding to constraints \eqref{eqn:constraint-node-selection} and \eqref{eqn:constraint-flow-conservation}, respectively. Hence, we have $A_1 \in \mathbb{R}^{L\cdot|V_P| \times (L+1)(|V_P|+|E_P|)}$ and $A_2 \in \mathbb{R}^{(L+1)\cdot|V_P| \times (L+1)(|V_P|+|E_P|)}$. The formulation \eqref{eqn:objective-single}-\eqref{eqn:constraint-binary-z} can be written as
\begin{align}
\begin{split}\label{eqn:formulation-matrix-form}
\min_y &\quad c_1^T y\\
\mbox{s.t.} &\quad A_1 y \leq 0\\
&\quad A_2 y = b_2\\
&\quad y \mbox{ binary vector}
\end{split}
\end{align}
where $c_1$ is a vector whose inner product with $y$ gives the service plus waiting time as in \eqref{eqn:objective-single}, and $b_2$ is a integral vector having $1$, $-1$ or $0$.

\begin{lemma}\label{lem:LP-relaxation-w-equality}
LP relaxation of the above formulation can be written as
\begin{align}
\begin{split}\label{eqn:LP-relaxation}
\min_y &\quad c^T x\\
\mbox{subject to} &\quad \begin{bmatrix}
A_1 & I_{12} & 0\\
A_2 & 0 & 0\\
I_{31} & 0 & I_{33}
\end{bmatrix} \begin{bmatrix}
y\\
s_1\\
s_2
\end{bmatrix} = \begin{bmatrix}
0\\
b_2\\
1
\end{bmatrix}\\
&\quad y, s_1, s_2 \geq 0
\end{split}
\end{align}
where $x$ is the vector $[y; s_1; s_2]$, and $s_1\in\mathbb{R}^{(L+1)(|V_P|+|E_P|)}$ and $s_2\in \mathbb{R}^{L|V_P|}$ are slack variables. The vector $c$ is the augmented vector of $c_1$ in \eqref{eqn:formulation-matrix-form} padded with zero to match the dimension of $x$. The matrices $I_{mn}$ are identity matrices of appropriate sizes.

\begin{IEEEproof}
%See Appendix \ref{app:proof-lemma-LP-relaxation-w-equality}.
To replace inequality constraint in \eqref{eqn:formulation-matrix-form} with equality constraint, we introduce the slack variable $s_1\geq 0$. The formulation \eqref{eqn:formulation-matrix-form} is equivalent to
\begin{align}
\begin{split}\label{eqn:formulation-matrix-form-eq}
\min &\quad c_1^T y\\
\mbox{s.t.} &\quad A_1 y + s_1 = 0\\
&\quad A_2 y = b_2\\
&\quad y \mbox{ binary vector}\\
&\quad s_1\geq 0
\end{split}
\end{align}
Now relax the binarity constraint as
\begin{align}
\begin{split}\label{eqn:formulation-matrix-form-eq-relaxed}
\min &\quad c_1^T y\\
\mbox{s.t.} &\quad A_1 y + s_1 = 0\\
&\quad A_2 y = b_2\\
&\quad y \leq 1\\
&\quad y, s_1\geq 0
\end{split}
\end{align}
The slack variable $s_2$ is introduced to replace the inequality constraint with equality constraint again, i.e., replace $y\leq 1$ with $y + s_2 = 1$ and $s_2\geq 0$. Putting all the equality constraints together obtains the desired result \eqref{eqn:LP-relaxation}.
\end{IEEEproof}
\end{lemma}

Note that the polyhedron in LP relaxation \eqref{eqn:LP-relaxation} is the same form as in Lemma \ref{lem:TU-to-integrality} with equality $Ax=b$, where
\begin{align}
A = \begin{bmatrix}
A_1 & I_{12} & 0\\
A_2 & 0 & 0\\
I_{31} & 0 & I_{33}
\end{bmatrix}, \quad x = \begin{bmatrix}
y\\ s_1\\ s_2
\end{bmatrix}, \quad b = \begin{bmatrix}
0\\ b_2 \\ 1
\end{bmatrix}
\end{align}
Note that $b$ is an integral vector. Consequently, by Lemmas \ref{lem:TU-to-integrality} and \ref{lem:LP-integrality}, we just need to show the total unimodularity of $A$.

\subsection{Total Unimodularity}
We start with some background on total unimodularity \cite{nemhauser:integer}. 
\begin{lemma}\label{lem:TU-equivalence}
For an $m\times n$ integral matrix $A$, the following are equivalent:
\begin{itemize}
\item[1)] $A$ is TU
\item[2)] $A^T$ is TU
\item[3)] For each $C\subseteq \{1,...,n\}$, there exists a partition $(C_1,C_2)$ of $C$ such that
\begin{align}
\left| \sum_{j\in C_1} a_{ij} - \sum_{j\in C_2} a_{ij}  \right| \leq 1 \mbox{ for } i=1,...,m \label{eqn:TU-partitioned-row-sum}
\end{align}
\item[4)] For each $R\subseteq \{1,...,m\}$, there exists a partition $(R_1,R_2)$ of $R$ such that
\begin{align}
\left| \sum_{i\in R_1} a_{ij} - \sum_{i\in R_2} a_{ij}  \right| \leq 1 \mbox{ for } j=1,...,j \label{eqn:TU-partitioned-col-sum}
\end{align}
\end{itemize}
\end{lemma}
Condition \eqref{eqn:TU-partitioned-row-sum} requires that for each row, the partitioned sum (called partitioned row sum) must differ by at most $1$. 
Note that statement 4) in Lemma \ref{lem:TU-equivalence} is an immediate consequence of 1), 2) and 3).

We can show the following lemma.
\begin{lemma}\label{lem:TU-A1A2-imply-TU-A}
If $[A_1; A_2]$ is TU, then $A$ is TU.

\begin{IEEEproof}
Suppose that $[A_1; A_2]$ is TU. Then, by statement 3) in Lemma \ref{lem:TU-equivalence}, any subset of the first block column (i.e., block containing $[A_1;A_2]$) can be partitioned into two sets $C_1$ and $C_2$ so that condition \eqref{eqn:TU-partitioned-row-sum} is satisfied. Note that the identity matrix $I_{31}$ in the first block column of $A$ does not affect the forming of $C_1$ and $C_2$ because any column partition of identity matrix satisfies condition \eqref{eqn:TU-partitioned-row-sum}. 

Next, consider adding an arbitrary column, say $i$th column, from the second block column of $A$. Note that the only nonzero entry (which is $1$) is in the $i$th row of this column. So, if the current partitioned sum difference of ith row is zero, then put the column into either $C_1$ or $C_2$, which does not violate condition \eqref{eqn:TU-partitioned-row-sum}. If the partitioned row sum difference is $1$, put the column into a subset with smaller row sum so that the partitioned row sum difference is balanced, i.e., zero. The third block column can be treated exactly the same as the second block in order to keep the condition in \eqref{eqn:TU-partitioned-row-sum}. This completes the proof.
\end{IEEEproof}
\end{lemma}

We can now focus on proving the total unimodularity of $[A_1; A_2]$ which is indeed totally unimodular.

\begin{lemma}\label{lem:TU-of-A1A2}
The matrix $[A_1; A_2]$ is TU.

\begin{IEEEproof}
Consider the breakdown of $[A_1; A_2]$ as shown in Fig.~\ref{fig:matrix-A}. Let $A_1 = [A_{11}\,\, A_{12}\,\, A_{13}]$. The first $|V_P|$ columns correspond to the variables $z_u,\forall u\in V_P$. The next $L|V_P|$ columns correspond to the variables $r_{u_{l-1}u_l}, \forall l=1,...,L, \forall u\in V_P$. The last $(L+1)|E_P|$ columns correspond to the variables $r_{uv}, \forall (u,v)\in E_l, l=0,1,...,L$.

Each row of $A_1$ in the top block corresponds to a constraint in \eqref{eqn:constraint-node-selection}. Constraints are arranged such that the first block of $L$ rows corresponds to some node $u$, and each row in the block indicates whether each cross-layer edge along replicated nodes of $u$ is visited. This is repeated for every other node in $V_P$. This forms the matrix $A_1$.

Each row of $A_2$ in the bottom block corresponds to a flow conservation constraint in \eqref{eqn:constraint-flow-conservation}. For instance, for some node $u\in V_P$, the first $L+1$ rows correspond to the flow conservation at replicated nodes $u_l$ in each layer $l=0,1,...,L$. As in $A_1$, the second block $A_{22}$ corresponds to the variables $r_{u_{l-1}u_l},l=1,...,L$ of cross-layer edges. Constraints are arranged in the order of $u_0,u_1,...,u_L$. Since the edges are between consecutive nodes, $A_{22}$ are written as in Fig. \ref{fig:matrix-A}. Note that this $A_{22}$ is a network matrix. The third block $A_{23}$ corresponds to the variables $r_{uv}, \forall (u,v)\in E_l, l=0,1,...,L$ of intra-layer edges. Clearly, $A_{23}$ is a network matrix of $L+1$ connected components $G_0,...,G_L$.

\begin{figure}[htbp]
\centerline{\includegraphics[scale=0.45]{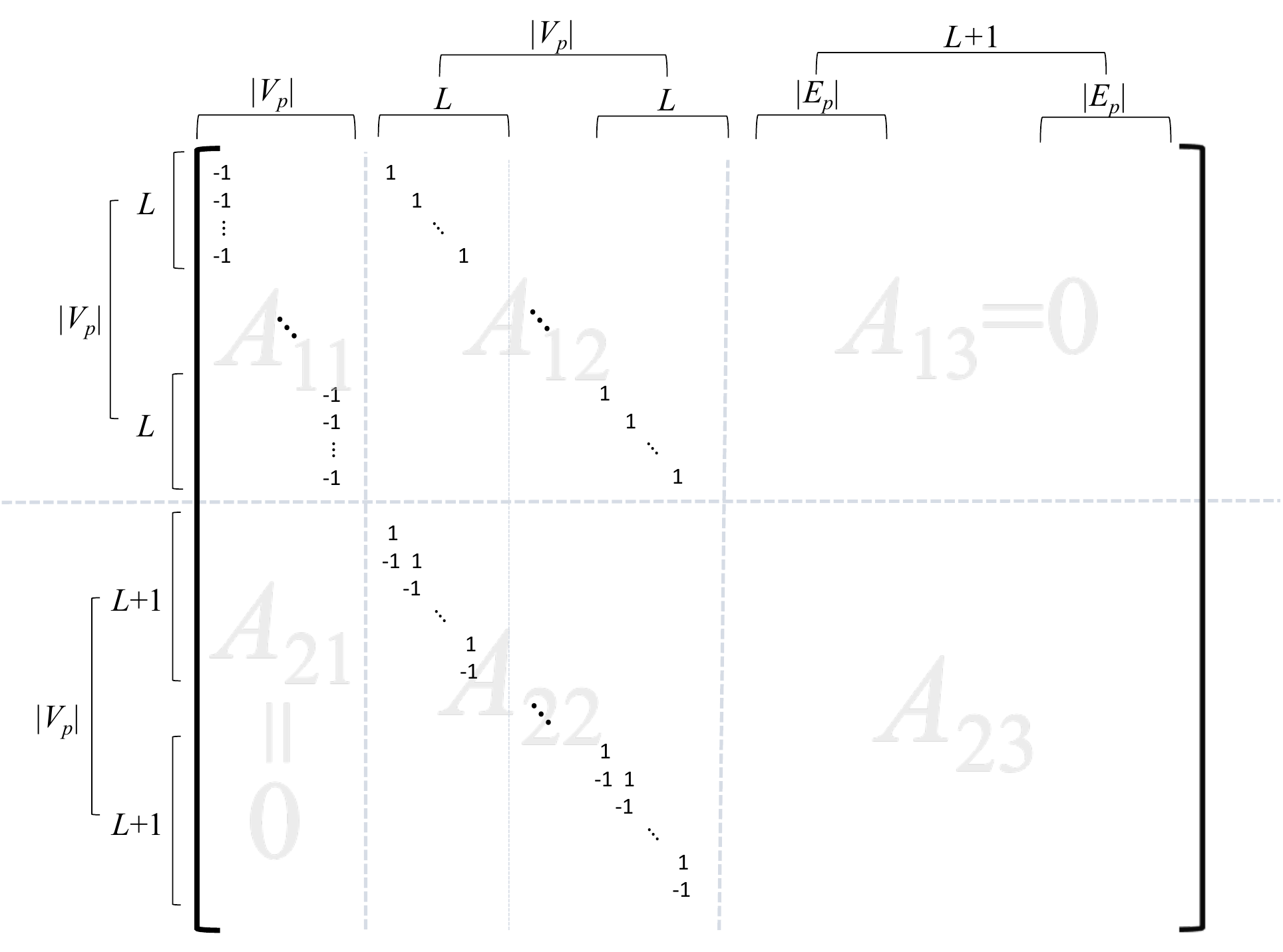}}
\caption{Breakdown of matrix $A$. $A_{23}$ is the network matrix of $L+1$ layers $G_0,...,G_L$ without cross-layer edges}
\label{fig:matrix-A}
\end{figure}

Consider partitioning the columns of the matrix $A$. Let $C_1$ and $C_2$ be a partition of an arbitrary subset $C$ of the entire columns. The following rule is applied for the partition.
\begin{itemize}
\item[1)] Put any column from the first $(L+1)|V_P|$ columns into $C_1$ regardless of $J$
\begin{itemize}
\item Without the third column block (i.e., the columns of $[A_{13}; A_{23}]$, this guarantees that each row sum is $0$, $1$, or $-1$. Furthermore, the row sums corresponding to $A_{22}$ have exactly the same numbers of $1$s and $-1$s. Let $s\in\mathbb{R}^{(L+1)|V_P|}$ be the vector of these row sums from the columns of $A_{22}$. Again, $s$ has exactly same numbers of $1$s and $-1$s. Consider the matrix $[s\,\, A_{23}]$. This matrix is obviously TU because i) $s$ has exactly the same numbers of $1$s and $-1$s and ii) each column of $A_{23}$ has exactly one $1$ and one $-1$. Consequently, any subset of rows can be partitioned into two subsets so that the condition in \eqref{eqn:TU-partitioned-col-sum} can be satisfied. That partition is indeed putting all the rows in the same subset, with which the partitioned column sum difference is always $0$, $1$, or $-1$. Consequently, $[s\,\, A_{23}]$ is TU and any subset of columns of the matrix can be partitioned so that the condition in \eqref{eqn:TU-partitioned-row-sum} is satisfied.
\end{itemize}
\item[2)] Treat any columns from the first two block columns as a single column by adding those columns. Let $d$ be the resulting column vector. Let $E$ be the matrix of arbitrary column(s) from $[A_{13}; A_{23}]$. Partition the columns of $[d\,\, E]$ in the same way as partitioning an arbitrary subset of columns of $[s, A23]$ discussed above.
\begin{itemize}
\item As shown in 1), all of the top block row sums satisfy condition \eqref{eqn:TU-partitioned-row-sum} since the third block column in the top block row (i.e., $A_{13}$) is a zero matrix. By the argument in 1) on the TU of $[s\,\, A_{23}]$, all of the bottom block row sums also satisfy \eqref{eqn:TU-partitioned-row-sum}.
\end{itemize}
\end{itemize}
This completes the proof.
\end{IEEEproof}
\end{lemma}

\textit{Proof of Theorem \ref{thm:greedy-analysis}}: By Lemmas \ref{lem:TU-of-A1A2} and \ref{lem:TU-A1A2-imply-TU-A}, $A$ is TU. By Lemmas \ref{lem:LP-integrality} and \ref{lem:TU-to-integrality}, LP relaxation \eqref{eqn:LP-relaxation} has an integral optimal solution. It is clear that LP relaxation \eqref{eqn:LP-relaxation} is equivalent to \eqref{eqn:LP-relaxation-short}. 
\begin{align}
\begin{split}\label{eqn:LP-relaxation-short}
\min &\quad c_1^T y\\
\mbox{s.t.} &\quad A_1 y \leq 0\\
&\quad A_2 y = b_2\\
&\quad 0\leq y \leq 1
\end{split}
\end{align}
Therefore, LP relaxation \eqref{eqn:LP-relaxation-short} has an integral optimal solution which is a single path with minimum completion time.

%\begin{theorem}\label{thm:LP-relaxation-works}
%By solving the following LP relaxation, a single path with minimum completion time can be found.
%\begin{align}
%\begin{split}\label{eqn:LP-relaxation-short}
%\min &\quad c_1^T y\\
%\mbox{s.t.} &\quad A_1 y \leq 0\\
%&\quad A_2 y = b_2\\
%&\quad 0\leq y \leq 1
%\end{split}
%\end{align}
%
%\begin{IEEEproof}
%By Lemmas \ref{lem:TU-of-A1A2} and \ref{lem:TU-A1A2-imply-TU-A}, $A$ is TU. By Lemmas \ref{lem:LP-integrality} and \ref{lem:TU-to-integrality}, LP relaxation \eqref{eqn:LP-relaxation} has an integral optimal solution. It is clear that LP relaxation \eqref{eqn:LP-relaxation} is equivalent to \eqref{eqn:LP-relaxation-short}. Therefore, LP relaxation \eqref{eqn:LP-relaxation-short} has an integral optimal solution which is a single path with minimum completion time.
%
%Note that if multiple fractional paths are found, it is straightforward to round the solution to a single unit path with no worse completion time. {\color{red} need to elaborate}
%\end{IEEEproof}
%\end{theorem}

\section{Proof of Theorem \ref{thm:greedy-analysis}}\label{app:proof-theorem-greedy-analysis}
We first introduce various routing policies below.
\begin{itemize}
\item Shortest completion (SC): optimal routing policy that achieves minimum job completion time $T^*$ in the actual system
\item Shortest service (SS): routing policy that achieves shortest service time for each job (same in the fictitious and actual systems)
\item Greedy (GR): Algorithm \ref{alg:greedy} (in the fictitious system)
\item Shortest waiting (SW): given that $p-1$ jobs are routed by greedy, the $p$th job is assigned to a node with positive computation capacity, say $u^*$, with shortest computation waiting time in the fictitious system. Then, a shortest transmission waiting (in the fictitious system) path is assigned from source to $u^*$ and from $u^*$ to destination.
\end{itemize}
The following notation is used:
\begin{itemize}
\item $W_j^\pi$: total waiting time of job $j$ under routing policy $\pi$
\item $S_j^\pi$: total service time of job $j$ under routing policy $\pi$
\item $W_j^{\pi}(tx)$ : total transmission waiting time of job $j$ under routing policy $\pi$
\item $W_j^{\pi}(cp)$ : total computation waiting time of job $j$ under routing policy $\pi$
\item $S_j^{\pi}(tx)$ : total transmission (service) time of job $j$ under routing policy $\pi$
\item $S_j^{\pi}(cp)$ : total computation (service) time of job $j$ under routing policy $\pi$
\end{itemize}
It is clear from the definition that $W_j^\pi + S_j^\pi$ is the completion time of job $j$ under routing policy $\pi$. We also have $S_j^\pi = S_j^\pi(tx) + S_j^\pi(cp)$ and $W_j^\pi = W_j^\pi(tx) + W_j^\pi(cp)$.

We need a few lemmas to prove Theorem \ref{thm:greedy-analysis}. For simplicity of presentation, we break the results into small lemmas and integrate these results for the proof of Theorem \ref{thm:greedy-analysis}. Denote by $j_p$ the $p$th job routed under greedy routing policy.
\begin{lemma}\label{lem:greedy-completion-time}
The job completion time under greedy algorithm in the actual system is upper-bounded by $W_{j_J}^{\scriptsize\mbox{GR}} + S_{j_J}^{\scriptsize\mbox{GR}}$ where $j_J$ is the last job routed under greedy algorithm.

\begin{IEEEproof}
By assumption of Theorem \ref{thm:greedy-analysis}, the greedy policy finds only a simple path for every job, and hence, each job's completion time in the fictitious system is the objective function value of the iteration in which the job's route is fixed. The greedy policy routes a job with earlier completion time first, and hence, the waiting time plus service time $W_{j_J}^{\scriptsize\mbox{GR}} + S_{j_J}^{\scriptsize\mbox{GR}}$ of last job routed under greedy is the job completion time of the greedy policy in the fictitious system. The completion time in the actual system is at most the completion time in the fictitious system, and consequently, upper-bounded by $W_{j_J}^{\scriptsize\mbox{GR}} + S_{j_J}^{\scriptsize\mbox{GR}}$.
\end{IEEEproof}
\end{lemma}

Consequently, in order to analyze the job completion time of greedy policy, we just need to derive a bound on $W_{j_J}^{\scriptsize\mbox{GR}} + S_{j_J}^{\scriptsize\mbox{GR}}$.

\begin{lemma}\label{lem:optimal-bounded-by-SS}
The optimal job completion time $T^*$ is lower-bounded as
\begin{align}
S_j^{\scriptsize\mbox{SS}} &\leq T^*, j=1,...,J\label{eqn:optimal-bounded-by-SS}\\
\frac{1}{|V_P| + |E_P|} \sum_{j=1}^{J} S_j^{\scriptsize\mbox{SS}} &\leq T^*\label{eqn:optimal-bounded-by-avgSS}
\end{align}

\begin{IEEEproof}
The LHS in inequality \eqref{eqn:optimal-bounded-by-SS} is the fastest possible service time of job $j$. The entire job is completed only after every job has been served. Hence, the minimum completion time $T^*$ cannot be smaller than the fastest possible service time of every job.

To show \eqref{eqn:optimal-bounded-by-avgSS}, we have
\begin{align}
\frac{1}{|V_P| + |E_P|} \sum_{j=1}^{J} S_j^{\scriptsize\mbox{SS}} \leq \frac{1}{|V_P| + |E_P|} \sum_{j=1}^{J} S_j^{\scriptsize\mbox{SC}} &\leq T^*
\end{align} 
The first inequality holds since the total service time of SC cannot be smaller than that of SS. In the second inequality, the LHS is the average busy time of network components (nodes+links) under SC. It is clear that job cannot be completed as long as there remains a busy component under SC, and consequently, optimal job completion time $T^*$ is no smaller than the LHS. Note that the service time at the link with infinite capacity is zero, and hence, those links should be excluded when computing the average. Similarly, under SS or SC, the computation will not be carried out at the node with zero computation capacity, and hence, those nodes should be excluded when computing the average. This completes the proof.
\end{IEEEproof}
\end{lemma}

The above lemma enables to compare GR and SC via SS. Let $h_L^j$ and $h_S^j$ be the longest and shortest path lengths (in $G_P$) in hop count between $s^j$ and $t^j$, respectively. Define $h_L=\max_j h_L^j$ and $h_S=\max_j h_S^j$.

\begin{lemma}\label{lem:servicetime-of-SW-bounded-by-SS}
The transmission times are bounded as
\begin{align}
S_j^{\scriptsize{\mbox{SW}}}(tx) &\leq 2\alpha_{tx} S_j^{\scriptsize\mbox{SS}}(tx), \forall j\label{eqn:servicetime-of-SW-bounded-by-SS}\\
S_j^{\scriptsize{\mbox{GR}}}(tx) &\leq (L+1)\alpha_{tx} S_j^{\scriptsize\mbox{SS}}(tx), \forall j\label{eqn:servicetime-of-GR-bounded-by-SS}
\end{align}
where $\alpha_{tx} = \frac{h_L\cdot\max\limits_{j,l} d_l^j \cdot \max\limits_{(u,v)\in E_P}\,\, \mu_{uv}}{h_S\cdot\min\limits_{j,l} d_l^j \cdot \min\limits_{(u,v)\in E_P} \,\,\mu_{uv}}$.

\begin{IEEEproof}
Under routing policy SW, link transmission occurs only in layers $0$ and $L$ because all the NN layers are computed in a single node with smallest computation waiting time. Furthermore, in layers $0$ and $L$, the transmission path is simple because SW takes a shortest transmission waiting path. Consequently, in both of layers $0$ and $L$, the transmission time is at most $ \frac{h_L\cdot\max\limits_{j,l} d_l^j}{\min\limits_{(u,v)\in E_P} \,\,\mu_{uv}}$. The transmission time under SW is thus upper-bounded as
\begin{align}
S_j^{\scriptsize{\mbox{SW}}}(tx) \leq 2 \frac{h_L\cdot\max\limits_{j,l} d_l^j}{\min\limits_{(u,v)\in E_P} \,\,\mu_{uv}}. \label{eqn:servicetime-of-SW-bound}
\end{align}

On the other hand, the transmission time under SS is at least $ \frac{h_S\cdot\min\limits_{j,l} d_l^j}{\max\limits_{(u,v)\in E_P} \,\,\mu_{uv}}$. Combining this with inequality in \eqref{eqn:servicetime-of-SW-bound} yields the desired result \eqref{eqn:servicetime-of-SW-bounded-by-SS}.

To prove \eqref{eqn:servicetime-of-GR-bounded-by-SS}, the same argument as above is applied to each of $L+1$ layers in $G$ since the greedy policy finds a simple path in each layer. This completes the proof.
\end{IEEEproof}
\end{lemma}

\begin{lemma}\label{lem:computationtime-bound-by-SS}
For any routing policy $\pi$, the computation service time is upper-bounded as
\begin{align}
S_j^{\pi}(cp) \leq \alpha_{cp} S_j^{\scriptsize\mbox{SS}}(cp), \forall j
\end{align}
where $\alpha_{cp} = \frac{\max\limits_u \mu_u}{\min\limits_{u} \mu_u}$.

\begin{IEEEproof}
The worst case is when all the layers are computed at the node with smallest (positive) computation capacity, and the best case is when computed at the node with largest computation capacity. The lemma immediately follows from this observation.
\end{IEEEproof}
\end{lemma}

\begin{lemma}\label{lem:servicetime-SW-bounded-by-SC}
The service time under routing policy SW is bounded as
\begin{align}
S_j^{\scriptsize\mbox{SW}} \leq \alpha_1 S_j^{\rm SS},\forall j,
\end{align}
where $\alpha_1 = \max(2\alpha_{tx}, \alpha_{cp})$.

\begin{IEEEproof}
We have
\begin{align}
S_j^{\scriptsize\mbox{SW}} &= S_j^{\scriptsize\mbox{SW}}(tx) + S_j^{\scriptsize\mbox{SW}}(cp)\\
&\leq 2\alpha_{tx} S_j^{\scriptsize\mbox{SS}}(tx) + \alpha_{cp} S_j^{\scriptsize\mbox{SS}}(cp)\\
&\leq \alpha_1 S_j^{\rm SS}
\end{align}
The first inequality follows from Lemmas \ref{lem:servicetime-of-SW-bounded-by-SS} and \ref{lem:computationtime-bound-by-SS}, and the second inequality follows from the definition of $\alpha_1$.
\end{IEEEproof}
\end{lemma}

The above lemmas characterize the bounds on the service time. We now derive the bounds on waiting time. Recall that given $p-1$ jobs routed by GR, the policy SW routes the $p$th job $j_p$ as mentioned in the beginning of this section.

\begin{lemma}\label{lem:tx-waiting-SW-bounded-by-SS}
Suppose that the original graph $G_P$ is $k$-edge-connected. Then,
\begin{align}
W_{j_J}^{\scriptsize{\mbox{SW}}}(tx) \leq \frac{2(L+1)\alpha_{tx}}{k} \sum_{p=1}^{J-1} S_{j_p}^{\scriptsize\mbox{SS}}(tx).
\end{align}

\begin{IEEEproof}
Recall that when job $j_J$ is to be routed by SW, the rest of $J-1$ jobs have already been routed by GR. By assumption, there are $k$ disjoint paths between any pair of nodes. For the path segment from source to $u^*$, the average transmission waiting time along each of $k$ disjoint paths is at most\footnote{This is an upper bound as there may edges not in $k$ disjoint paths.} $\frac{1}{k}\sum\limits_{p=1}^{J-1} S_{j_p}^{\scriptsize\mbox{GR}}(tx)$. This shows that there is a path from source to $u^*$ with transmission waiting time no greater than $\frac{1}{k}\sum\limits_{p=1}^{J-1} S_{j_p}^{\scriptsize\mbox{GR}}(tx)$. Since the routing policy SW selects a path from source to $u^*$ that has the smallest transmission waiting time, the transmission waiting time $W_{j_J}^{\scriptsize{\mbox{SW}}}(tx)$ is upper-bounded by $\frac{1}{k}\sum\limits_{p=1}^{J-1} S_{j_p}^{\scriptsize\mbox{GR}}(tx)$. By Lemma \ref{lem:servicetime-of-SW-bounded-by-SS}, this is in turn upper-bounded by $\frac{(L+1)}{k}\sum\limits_{p=1}^{J-1} S_{j_p}^{\scriptsize\mbox{SS}}(tx)$. Applying the same argument to the segment from $u^*$ to destination proves the lemma.
\end{IEEEproof}
\end{lemma}

\begin{lemma}\label{lem:cp-waiting-SW-bounded-by-SS}
We have
\begin{align}
W_{j_J}^{\scriptsize\mbox{SW}}(cp) \leq \frac{\alpha_{cp}}{|V_P|} \sum_{p=1}^{J-1} S_{j_p}^{\scriptsize\mbox{SS}}(cp)
\end{align}

\begin{IEEEproof}
This lemma immediately follows from the following inequalities:
\begin{align}
W_{j_J}^{\scriptsize\mbox{SW}}(cp) \leq \frac{1}{|V_P|} \sum_{p=1}^{J-1} S_{j_p}^{\scriptsize\mbox{GR}}(cp) \leq \frac{\alpha_{cp}}{|V_P|} \sum_{p=1}^{J-1} S_{j_p}^{\scriptsize\mbox{SS}}(cp)
\end{align}
Recall that both GR (by nature of formulation \eqref{eqn:objective-single}-\eqref{eqn:constraint-binary-z}) and SW (by definition) do not compute at the node with zero computation capacity. In addition, $|V_P|$ is the number of nodes with positive computation capacity. The RHS of the first inequality is thus the average computation waiting time at a node.  Since the routing policy SW selects a node with minimum waiting time, the node waiting time under SW should be no greater than the average node waiting time, which shows the first inequality. The second inequality follows from Lemma \ref{lem:computationtime-bound-by-SS}. This completes the proof.
\end{IEEEproof}
\end{lemma}

\begin{lemma}\label{lem:GR-bounded-by-SW}
We have
\begin{align}
W_{j_J}^{\scriptsize\mbox{GR}} + S_{j_J}^{\scriptsize\mbox{GR}} \leq W_{j_J}^{\scriptsize\mbox{SW}} + S_{j_J}^{\scriptsize\mbox{SW}}.
\end{align}

\begin{IEEEproof}
Recall that given $J-1$ jobs routed by GR, SW finds a path such that the last job is processed at a single node, say $u^*$, with minimum waiting time, and a shortest (in link transmission waiting) path from source to $u^*$ and a shortest path from $u^*$ to destination are concatenated to form a path from source to destination of the last job. On the other hand, for the last job, GR finds a path with minimum waiting plus service time, given $J-1$ jobs routed by GR. Therefore, the inequality holds.
\end{IEEEproof}
\end{lemma}

\textit{Proof of Theorem \ref{thm:greedy-analysis}}: By Lemmas \ref{lem:greedy-completion-time} and \ref{lem:GR-bounded-by-SW}, the job completion time under GR in the actual system is at most $W_{j_J}^{\scriptsize\mbox{SW}} + S_{j_J}^{\scriptsize\mbox{SW}}$. Let $\alpha_2 = \max\left(\frac{\alpha_{cp}}{|V_P|}, \frac{2(L+1)\alpha_{tx}}{k}\right)$. By Lemmas \ref{lem:tx-waiting-SW-bounded-by-SS} and \ref{lem:cp-waiting-SW-bounded-by-SS}, we have
\begin{align}
W_{j_J}^{\scriptsize{\mbox{SW}}} &\leq \frac{\alpha_{cp}}{|V_P|} \sum_{p=1}^{J-1} S_{j_p}^{\scriptsize\mbox{SS}}(cp) + \frac{2(L+1)\alpha_{tx}}{k}\sum_{p=1}^{J-1} S_{j_p}^{\scriptsize\mbox{SS}}(tx)\\
&\leq \alpha_2 \sum_{p=1}^{J-1} S_{j_p}^{\scriptsize\mbox{SS}}
\end{align}
This together with Lemma \ref{lem:servicetime-SW-bounded-by-SC} leads to
{\small
\begin{align}
&W_{j_J}^{\scriptsize\mbox{SW}} + S_{j_J}^{\scriptsize\mbox{SW}} \leq \alpha_2 \sum_{p=1}^{J-1} S_{j_p}^{\scriptsize\mbox{SS}} + \alpha_1 S_{j_J}^{\scriptsize\mbox{SS}}\\
&\leq \alpha_3\left\{\frac{1}{|V_P|+|E_P|}\sum_{p=1}^{J-1} S_{j_p}^{\scriptsize\mbox{SS}} +  S_{j_J}^{\scriptsize\mbox{SS}}\right\}\\
&= \alpha_3\left\{\frac{1}{|V_P|+|E_P|}\sum_{p=1}^{J} S_{j_p}^{\scriptsize\mbox{SS}} + \left(1 - \frac{1}{|V_P|+|E_P|}\right) S_{j_J}^{\scriptsize\mbox{SS}}\right\}\\
&\leq \alpha_3 \left(2 - \frac{1}{|V_P|+|E_P|}\right)T^*
\end{align}
}
where $\alpha_3 = \max(\alpha_2(|V_P|+|E_P|), \alpha_1)$. The last inequality follows from Lemma \ref{lem:optimal-bounded-by-SS}. This completes the proof.

\section{Proof of Corollary \ref{cor:two-approx}}\label{app:proof-corollary-2-approx}
By assumption of zero network delay, we have $|E_P|=0$, and also $\alpha_{tx}=0$ because Lemma \ref{lem:servicetime-of-SW-bounded-by-SS} holds with $\alpha_{tx}=0$. In addition, $\alpha_{cp}=1$. It immediately follows that $\alpha_1=1$, $\alpha_2=\frac{1}{|V_P|}$ and $\alpha_3=1$. Therefore, the approximation ratio is $2-\frac{1}{|V_P|}$, which completes the proof.

%It now follows that
%{\small
%\begin{align}
%W_J^{\scriptsize\mbox{SW}} + S_J^{\scriptsize\mbox{SW}} &\leq \frac{\alpha_{cp}}{|V_P|} \sum_{j=1}^{J-1} S_j^{\scriptsize\mbox{SS}}(cp) + \frac{2(L+1)}{k}\sum_{j=1}^{J-1} S_j^{\scriptsize\mbox{SS}}(tx) + 
%\alpha_1 S_{j_J}^{\scriptsize\mbox{SS}}\\
%&\leq (|V_P|+|E_P|)\cdot \alpha_2 \frac{1}{|V_P|+|E_P|} \sum_{j=1}^{J-1} S_j^{\scriptsize\mbox{SS}} + \alpha_1 T^*\\
%&\leq (|V_P|+|E_P|)\cdot\max\left(\frac{\alpha_{cp}}{|V_P|}, \frac{2(L+1)\alpha_{tx}}{k}\right) (1 - \frac{1}{|V_P|+|E_P|}) T^* + \alpha_1 T^*\\
%&= \left\{(|V_P|+|E_P| -1)\cdot\max\left(\frac{\alpha_{cp}}{|V_P|}, \frac{2(L+1)\alpha_{tx}}{k}\right) + \alpha_1\right\} T^*
%\end{align}}
%\end{comment}

\section{Formulation of Optimal Routing}\label{app:opt-algo}
Consider the binary variable $t^j_p$ that takes the value 1 if job $j$ has priority $p$, and 0 otherwise. For two jobs $j$ and $k$, if $t^j_p=1$ and $t^k_{p'}=1$ with $p'<p$, then it means that job $k$ has higher priority than $j$. As mentioned in Section~\ref{sec:applications}, the waiting time of $j$ at a component (link or node) in the fictitious system is the sum of the computation or transmission tasks of all the jobs passing through the component with higher priority than $j$. The problem of minimizing the completion time in the fictitious system can be formulated as follows:
{\small
\begin{align}
\min_{r,z,t} &\,\, \max_{j\in\mathcal{J}} \sum_{(u,v)\in E} \frac{q^j_{uv}}{\mu_{uv}} r^j_{uv} + \nonumber\\
&\,\, \sum_{(u,v)\in E\setminus E_C} \frac{1}{\mu_{uv}} \left( Q_{uv} +\sum_{p\in\mathcal{J}}\sum_{p'<p} \sum_{j'\in\mathcal{J}} t^j_p t^{j'}_{p'} q^{j'}_{uv} r^{j'}_{uv} \right) r^j_{uv} + \nonumber\\
&\,\,\sum_{u\in V_P}\frac{1}{\mu_u} \left(  Q_u + \sum_{\substack{(v,w)\in E_C\\v\in V_u}}\sum_{p\in\mathcal{J}}\sum_{p'<p} \sum_{j'\in\mathcal{J}} t^j_p t^{j'}_{p'} q^{j'}_{vw} r^{j'}_{vw} \right)z^j_u\label{eqn:objective-single-app}\\
\mbox{s.t.} &\,\, \mbox{Constraint \eqref{eqn:con-path}}\nonumber\\
& \,\, z^j_u\geq r^j_{u_{l-1}u_l},\forall u\in V_P, l=1,...,L^j, j\in\mathcal{J}\label{eqn:constraint-node-selection-app}\\
&\,\, \sum_{p\in\mathcal{J}}t^j_p = 1,\forall j\in \mathcal{J}\label{eqn:constraint-priority-select-j-app}\\
&\,\, \sum_{j\in\mathcal{J}}t^j_p = 1,\forall p\in \mathcal{J}\label{eqn:constraint-priority-select-p-app}\\
&\,\, r^j_{uv}\in\{0,1\},\forall (u,v)\in E, \forall j\in \mathcal{J} \label{eqn:constraint-binary-r-app}\\
&\,\, z^j_{u}\in\{0,1\},\forall u\in V_P, \forall j\in \mathcal{J} \label{eqn:constraint-binary-z-app}\\
&\,\, t_{jp}\in\{0,1\},\forall j,p\in \mathcal{J} \label{eqn:constraint-binary-t-app}
\end{align}
}
The first term in the objective function is the total service time. The second term is the waiting time at the traversing links, adding up the transmission tasks of all the jobs with higher priority. Similarly, the third term is the waiting time at the node where the processing of job $j$ is carried out, adding up the computation tasks of all the jobs with higher priority. The constraints \eqref{eqn:constraint-priority-select-j-app} and \eqref{eqn:constraint-priority-select-p-app} ensure that every job is assigned a unique priority index $p$, and no priority index $p$ is assigned to more than one job. The maximum of this objective function is the completion time, and the goal is to find a routing with minimum completion time. This formulation introduces a large number of variables. We found that it takes excessively long time to preprocess for constructing objective function and constraints. Furthermore, the solver often fails to find a solution even in about half an hour. This led us to exclude problem instances with large network/job size.

\section{Node-first Selection Algorithm}\label{app:nfs-algo}
Consider the routine $[u, C] = \mathsf{selectNode}(V, W)$ that selects the smallest-weight node in $V$ with node weights $W=[w_v, \forall v\in V]$. The output $u$ is the selected node, and $C$ is the weight of the selected node. Likewise, consider the routine $[H, C] = \mathsf{findShoPath}(G, u, v, W)$ that finds the shortest path from $u\in V$ to $v\in V$ over the graph $G=(V,E)$ with link weights $W=[w_{uv},\forall (u,v)\in E]$. The output $H$ is the shortest path found, and $C$ is the total weight of the shortest path. Consider the following algorithm:

\begin{algorithm}%[H]
\SetAlgoLined
\KwGiven{Jobs $\mathcal{J}=\{1,...,J\}$
}
\KwInit{$Q_u=0,\forall u\in V_P$; $Q_{uv}=0,\forall (u,v)\in E_P$; $U = \mathcal{J}$; $p=1;$}
\While{$U \neq \emptyset$}{\small
$W_{cp}^j = \left[\frac{Q_u + \sum_{l=1}^{L^j}c^j_l}{\mu_u}, \forall u\in V_P\right],\forall j\in U$\;
$[u^j, C^j_{cp}] = \mathsf{selectNode}(V_P, W_{cp}^j),\forall j\in U$\;
$W_{tx}^j = \left[\frac{Q_{uv}+d^j_0}{\mu_{uv}}, \forall (u,v)\in E_P\right],\forall j\in U$\;
$[H^j_1, C^j_{tx,1}] = \mathsf{findShoPath}(G_P, s^j, u^j, W_{tx}^j),\forall j\in U$\;
$W_{tx}^j = \left[\frac{Q_{uv}+d^j_{L^j}}{\mu_{uv}}, \forall (u,v)\in E_P\right],\forall j\in U$\;
$[H^j_2, C^j_{tx,2}] = \mathsf{findShoPath}(G_P, u^j, t^j, W_{tx}^j),\forall j\in U$\;
$j_p = \arg\min\limits_{j\in U}  C^j_{tx,1} + C^j_{cp} +  C^j_{tx,2}$\;
$Q_{u^j} \leftarrow Q_{u^j} + \sum_{l=1}^{L^{j_p}} c_l^{j_p}$\;
$Q_{uv} \leftarrow Q_{uv} + d_0^{j_p} ,\forall (u,v)\in H^{j_p}_1$\;
$Q_{uv} \leftarrow Q_{uv} + d_{L^j}^{j_p} ,\forall (u,v)\in H^{j_p}_2$\;
$p \leftarrow p+1$\;
$U \leftarrow U\setminus \{j_p\}$\;
}
\KwOutput{ Priority \& Routing: $[j_1>\cdots>j_J]$ \& $[H^{j_p}_1\rightarrow H^{j_p}_2, \forall p=1,...,J]$\
%\KwOutput{ Priority: $j_1>\cdots>j_J$\\
%\hspace{1.45cm}Routing: $r_{uv}^*(j_p), \forall (u,v), p=1,...,J$\
}
\caption{Node-first Selection Algorithm}\label{alg:nfs}
\end{algorithm}

In line 2, $W_{cp}^j$ is the vector of computation completion time when the entire layer of job $j$ is assigned to each node. In line 3, for each job, the node with shortest computation completion time is selected together with the corresponding completion time. In line 4, the weight vector $W_{tx}^j$ contains the transmission completion time at each link if the input data of each job is transferred over the link. In line 5, the path from source to the best node (selected in line 3) found such that the input data are delivered to the best node with minimum transmission latency. Similarly, in lines 6-7, the path from the best node to the destination is found with respect to the output data, i.e., the shortest path from the best node to the destination in terms of transmission latency. In line 8, the job with the earliest completion selected assuming that each job is computed at the node selected in line 3, and the input and output data are delivered along the paths found in lines 5 and 7, respectively. The route of the selected job is fixed with priority corresponding to $p$ (the lower, the higher priority), and the unfinished tasks $Q$ are updated based on the fixed path of the job. In addition, the selected job is removed from the set $U$ that contains unassigned jobs. This is repeated until all the jobs are assigned. Therefore, the output of this algorithm gives the priority and the path (with single node selection) for every job. 

This algorithm puts an emphasis on the computation by first selecting the node with earliest completion if the entire layer of a job is assigned to a single node. Although the algorithm may be able to give out a solution quickly, there are two drawbacks. First, it assigns the entire layer to a single node, and hence, in situations where some layers need to be split, the algorithm may perform poor. Second, in the setting where data rates are low, the transmission time may become a substantial element in the completion time. In this case, the node-first selection strategy may incur large transmission latency, eventually leading to large completion time.

\bibliographystyle{IEEEtran}
\bibliography{reference}

% Generated by IEEEtran.bst, version: 1.12 (2007/01/11)
\begin{thebibliography}{10}
\providecommand{\url}[1]{#1}
\csname url@samestyle\endcsname
\providecommand{\newblock}{\relax}
\providecommand{\bibinfo}[2]{#2}
\providecommand{\BIBentrySTDinterwordspacing}{\spaceskip=0pt\relax}
\providecommand{\BIBentryALTinterwordstretchfactor}{4}
\providecommand{\BIBentryALTinterwordspacing}{\spaceskip=\fontdimen2\font plus
\BIBentryALTinterwordstretchfactor\fontdimen3\font minus
  \fontdimen4\font\relax}
\providecommand{\BIBforeignlanguage}[2]{{%
\expandafter\ifx\csname l@#1\endcsname\relax
\typeout{** WARNING: IEEEtran.bst: No hyphenation pattern has been}%
\typeout{** loaded for the language `#1'. Using the pattern for}%
\typeout{** the default language instead.}%
\else
\language=\csname l@#1\endcsname
\fi
#2}}
\providecommand{\BIBdecl}{\relax}
\BIBdecl

\bibitem{letaief:roadmap}
K.~B. Letaief, W.~Chen, Y.~Shi, J.~Zhang, and Y.-J.~A. Zhang, ``{The Roadmap to
  6G: AI Empowered Wireless Networks},'' \emph{IEEE Communications Magazine},
  vol.~57, no.~8, pp. 84--90, 2019.

\bibitem{DBLP:journals/corr/IandolaMAHDK16}
\BIBentryALTinterwordspacing
F.~N. Iandola, M.~W. Moskewicz, K.~Ashraf, S.~Han, W.~J. Dally, and K.~Keutzer,
  ``{SqueezeNet: AlexNet-level accuracy with 50x fewer parameters and
  {\textless}1MB model size},'' \emph{CoRR}, vol. abs/1602.07360, 2016.
  [Online]. Available: \url{http://arxiv.org/abs/1602.07360}
\BIBentrySTDinterwordspacing

\bibitem{DBLP:journals/corr/HowardZCKWWAA17}
\BIBentryALTinterwordspacing
A.~G. Howard, M.~Zhu, B.~Chen, D.~Kalenichenko, W.~Wang, T.~Weyand,
  M.~Andreetto, and H.~Adam, ``{MobileNets: Efficient Convolutional Neural
  Networks for Mobile Vision Applications},'' \emph{CoRR}, vol. abs/1704.04861,
  2017. [Online]. Available: \url{http://arxiv.org/abs/1704.04861}
\BIBentrySTDinterwordspacing

\bibitem{howard:mobilenetv3}
A.~Howard, M.~Sandler, B.~Chen, W.~Wang, L.-C. Chen, M.~Tan, G.~Chu,
  V.~Vasudevan, Y.~Zhu, R.~Pang, H.~Adam, and Q.~Le, ``Searching for
  mobilenetv3,'' in \emph{2019 IEEE/CVF International Conference on Computer
  Vision (ICCV)}, 2019, pp. 1314--1324.

\bibitem{ren:survey}
\BIBentryALTinterwordspacing
P.~Ren, Y.~Xiao, X.~Chang, P.-y. Huang, Z.~Li, X.~Chen, and X.~Wang, ``A
  comprehensive survey of neural architecture search: Challenges and
  solutions,'' vol.~54, no.~4, 2021. [Online]. Available:
  \url{https://doi.org/10.1145/3447582}
\BIBentrySTDinterwordspacing

\bibitem{baccour:RLPDNN}
E.~Baccour, A.~Erbad, A.~Mohamed, M.~Hamdi, and M.~Guizani, ``{RL-PDNN:
  Reinforcement Learning for Privacy-Aware Distributed Neural Networks in IoT
  Systems},'' \emph{IEEE Access}, vol.~9, pp. 54\,872--54\,887, 2021.

\bibitem{disabato:DDCNN}
S.~Disabato, M.~Roveri, and C.~Alippi, ``{Distributed Deep Convolutional Neural
  Networks for the Internet-of-Things},'' \emph{IEEE Transactions on
  Computers}, vol.~70, no.~8, pp. 1239--1252, 2021.

\bibitem{shi:edge}
W.~Shi, J.~Cao, Q.~Zhang, Y.~Li, and L.~Xu, ``Edge computing: Vision and
  challenges,'' \emph{IEEE Internet of Things Journal}, vol.~3, no.~5, pp.
  637--646, 2016.

\bibitem{wang:convergence}
X.~Wang, Y.~Han, V.~C.~M. Leung, D.~Niyato, X.~Yan, and X.~Chen, ``{Convergence
  of Edge Computing and Deep Learning: A Comprehensive Survey},'' \emph{IEEE
  Communications Surveys Tutorials}, vol.~22, no.~2, pp. 869--904, 2020.

\bibitem{liu:dare}
Q.~Liu and T.~Han, ``{DARE: Dynamic Adaptive Mobile Augmented Reality with Edge
  Computing},'' in \emph{ICNP}, 2018, pp. 1--11.

\bibitem{tan:deep}
T.~Tan and G.~Cao, ``Deep learning video analytics through edge computing and
  neural processing units on mobile devices,'' \emph{IEEE Transactions on
  Mobile Computing}, pp. 1--1, 2021.

\bibitem{wu:accuracy}
W.~Wu, P.~Yang, W.~Zhang, C.~Zhou, and X.~Shen, ``Accuracy-guaranteed
  collaborative dnn inference in industrial iot via deep reinforcement
  learning,'' \emph{IEEE Transactions on Industrial Informatics}, vol.~17,
  no.~7, pp. 4988--4998, 2021.

\bibitem{pascale:network}
E.~Di~Pascale, I.~Macaluso, A.~Nag, M.~Kelly, and L.~Doyle, ``{The Network As a
  Computer: A Framework for Distributed Computing Over IoT Mesh Networks},''
  \emph{IEEE Internet of Things Journal}, vol.~5, no.~3, pp. 2107--2119, 2018.

\bibitem{zhao:deepthings}
Z.~Zhao, K.~M. Barijough, and A.~Gerstlauer, ``{DeepThings: Distributed
  Adaptive Deep Learning Inference on Resource-Constrained IoT Edge
  Clusters},'' \emph{IEEE Transactions on Computer-Aided Design of Integrated
  Circuits and Systems}, vol.~37, no.~11, pp. 2348--2359, 2018.

\bibitem{mao:modnn}
J.~Mao, X.~Chen, K.~W. Nixon, C.~Krieger, and Y.~Chen, ``{MoDNN: Local
  distributed mobile computing system for Deep Neural Network},'' in
  \emph{Design, Automation Test in Europe Conference Exhibition (DATE), 2017},
  2017, pp. 1396--1401.

\bibitem{hadidi:toward}
R.~Hadidi, J.~Cao, M.~S. Ryoo, and H.~Kim, ``Toward collaborative inferencing
  of deep neural networks on internet-of-things devices,'' \emph{IEEE Internet
  of Things Journal}, vol.~7, no.~6, pp. 4950--4960, 2020.

\bibitem{stahl:fully}
R.~Stahl, Z.~Zhao, D.~Mueller-Gritschneder, A.~Gerstlauer, and U.~Schlichtmann,
  ``Fully distributed deep learning inference on resource-constrained edge
  devices,'' in \emph{International Conference on Embedded Computer Systems:
  Architectures, Modeling, and Simulation}, D.~N. Pnevmatikatos, M.~Pelcat, and
  M.~Jung, Eds., 2019, pp. 77--90.

\bibitem{xue:edgeld}
F.~Xue, W.~Fang, W.~Xu, Q.~Wang, X.~Ma, and Y.~Ding, ``{EdgeLD: Locally
  Distributed Deep Learning Inference on Edge Device Clusters},'' in \emph{IEEE
  International Conference on High Performance Computing and Communications},
  2020, pp. 613--619.

\bibitem{chang:EDDL}
Y.~Chang, X.~Huang, Z.~Shao, and Y.~Yang, ``{An Efficient Distributed Deep
  Learning Framework for Fog-Based IoT Systems},'' in \emph{2019 IEEE Global
  Communications Conference (GLOBECOM)}, 2019, pp. 1--6.

\bibitem{he:joint}
W.~He, S.~Guo, S.~Guo, X.~Qiu, and F.~Qi, ``Joint dnn partition deployment and
  resource allocation for delay-sensitive deep learning inference in iot,''
  \emph{IEEE Internet of Things Journal}, vol.~7, no.~10, pp. 9241--9254, 2020.

\bibitem{zhang:learning}
C.~Zhang, W.~Song, Z.~Cao, J.~Zhang, P.~S. Tan, and X.~Chi, ``Learning to
  dispatch for job shop scheduling via deep reinforcement learning,'' in
  \emph{Advances in Neural Information Processing Systems}, vol.~33, 2020, pp.
  1621--1632.

\bibitem{huang:toward}
Y.~Huang, X.~Qiao, S.~Dustdar, J.~Zhang, and J.~Li, ``Toward decentralized and
  collaborative deep learning inference for intelligent iot devices,''
  \emph{IEEE Network}, vol.~36, no.~1, pp. 59--68, 2022.

\bibitem{huang:enabling}
Y.~Huang, X.~Qiao, W.~Lai, S.~Dustdar, J.~Zhang, and J.~Li, ``Enabling dnn
  acceleration with data and model parallelization over ubiquitous end
  devices,'' \emph{IEEE Internet of Things Journal}, pp. 1--1, 2021.

\bibitem{xue:ddpqn}
M.~Xue, H.~Wu, G.~Peng, and K.~Wolter, ``Ddpqn: An efficient dnn offloading
  strategy in local-edge-cloud collaborative environments,'' \emph{IEEE
  Transactions on Services Computing}, vol.~15, no.~2, pp. 640--655, 2022.

\bibitem{xu:energy}
Z.~Xu, L.~Zhao, W.~Liang, O.~F. Rana, P.~Zhou, Q.~Xia, W.~Xu, and G.~Wu,
  ``Energy-aware inference offloading for dnn-driven applications in mobile
  edge clouds,'' \emph{IEEE Transactions on Parallel and Distributed Systems},
  vol.~32, no.~4, pp. 799--814, 2021.

\bibitem{garey:computers}
M.~R. Garey and D.~S. Johnson, \emph{Computers and Intractability; A Guide to
  the Theory of NP-Completeness}.\hskip 1em plus 0.5em minus 0.4em\relax USA:
  W. H. Freeman \& Co., 1990.

\bibitem{cao:enhancing}
Z.~Cao, S.~S. Panwar, M.~Kodialam, and T.~V. Lakshman, ``Enhancing mobile
  networks with software defined networking and cloud computing,''
  \emph{IEEE/ACM Trans. on Netw.}, vol.~25, no.~3, pp. 1431--1444, 2017.

\bibitem{zhang:optimal}
J.~Zhang, A.~Sinha, J.~Llorca, A.~M. Tulino, and E.~Modiano, ``Optimal control
  of distributed computing networks with mixed-cast traffic flows,''
  \emph{IEEE/ACM Trans. on Networking}, vol.~29, no.~4, pp. 1760--1773, 2021.

\bibitem{wang:energy}
Q.~Wang, J.~Fu, J.~Wu, B.~Moran, and M.~Zukerman, ``Energy-efficient
  priority-based scheduling for wireless network slicing,'' in \emph{2018 IEEE
  Global Communications Conference (GLOBECOM)}, 2018, pp. 1--6.

\bibitem{jiang:network}
M.~Jiang, M.~Condoluci, and T.~Mahmoodi, ``Network slicing management amp;
  prioritization in 5g mobile systems,'' in \emph{22th European Wireless
  Conference}, 2016, pp. 1--6.

\bibitem{nemhauser:integer}
G.~L. Nemhauser and L.~A. Wolsey, \emph{Integer and Combinatorial
  Optimization}.\hskip 1em plus 0.5em minus 0.4em\relax New York, NY, USA:
  Wiley-Interscience, 1999.

\end{thebibliography}

%\newpage
%
%\section{Biography Section}
%If you have an EPS/PDF photo (graphicx package needed), extra braces are
% needed around the contents of the optional argument to biography to prevent
% the LaTeX parser from getting confused when it sees the complicated
% $\backslash${\tt{includegraphics}} command within an optional argument. (You can create
% your own custom macro containing the $\backslash${\tt{includegraphics}} command to make things
% simpler here.)
% 
%\vspace{11pt}

%\bf{If you include a photo:}\vspace{-33pt}
\begin{IEEEbiography}{Sehun Jung}
is an Undergraduate Student in the Department of Computer Science and Engineering, Konkuk University, Seoul, South Korea. His research interests include reinforcement learning and its applications to neural network training and computation.
\end{IEEEbiography}

\begin{IEEEbiography}{Hyang-Won Lee}
(S’02-M’08) received the B.S., M.S., and Ph.D. degrees all in Electrical Engineering and Computer Science from the Korea Advanced
Institute of Science and Technology, Daejeon, South Korea, in 2001, 2003, and 2007, respectively. He was a Post-Doctoral Research Associate with the Massachusetts Institute of Technology (MIT) from 2007 to 2011, and a Research Assistant Professor with KAIST from 2011 and 2012. He was also a Visiting Research Scientist at the Laboratory of Information and Decision Systems (LIDS), MIT from 2017 to 2018. He is presently a Professor with the Department of Computer Science and Engineering, Konkuk University, Seoul, South Korea. His research interests are in the areas of network optimization and algorithms, reinforcement learning and formal verification.
\end{IEEEbiography}

%\bf{If you include a photo:}\vspace{-33pt}
%\begin{IEEEbiography}[{\includegraphics[width=1in,height=1.25in,clip,keepaspectratio]{fig1}}]{Hyang-Won Lee}
%Use $\backslash${\tt{begin\{IEEEbiography\}}} and then for the 1st argument use $\backslash${\tt{includegraphics}} to declare and link the author photo.
%Use the author name as the 3rd argument followed by the biography text.
%\end{IEEEbiography}
%\vspace{11pt}
%
%%\bf{If you will not include a photo:}\vspace{-33pt}
%\begin{IEEEbiographynophoto}{John Doe}
%Use $\backslash${\tt{begin\{IEEEbiographynophoto\}}} and the author name as the argument followed by the biography text.
%\end{IEEEbiographynophoto}

\vfill

\end{document}